\documentclass[preprint,12pt]{elsarticle}

\usepackage{graphicx}
\usepackage{dcolumn}
\usepackage{bm}
\usepackage{amssymb}
\usepackage{hyperref}
\usepackage{amsmath}
\usepackage{fancyhdr}
\usepackage[normalem]{ulem}
\usepackage{amsmath}
\usepackage{textcomp, gensymb}
\usepackage{url}
\usepackage{amsmath,amssymb,stmaryrd}   
\usepackage{listings}
\usepackage{tcolorbox}
\usepackage{blindtext}  
\usepackage{tikz,lipsum,lmodern}
\usepackage{xcolor}     
\usepackage{lipsum}     
\usepackage{ntheorem}   
\usepackage{mdframed}   
\usepackage{gensymb}
\usepackage{setspace}
\usepackage{stackrel}
\usepackage{tikz-cd}
\usepackage{algorithm}
\usepackage{algpseudocode}

\journal{Computational Mechanics}
\usepackage[english]{babel}
\usepackage[intoc, english]{nomencl}
\makenomenclature
\begin{document}

\begin{frontmatter}



\title{Stress Predictions in Polycrystal Plasticity using Graph Neural Networks with Subgraph Training}


\author[inst1]{Hanfeng Zhai\footnote{E-mail: \texttt{hzhai@stanford.edu}}}

\affiliation[inst1]{organization={Department of Mechanical Engineering,\\ Stanford University},
            city={Stanford},
            postcode={94305}, 
            state={CA},
            country={USA}}


\begin{abstract}\small
Numerical modeling of polycrystal plasticity is computationally intensive. We employ Graph Neural Networks (GNN) to predict stresses on complex geometries for polycrystal plasticity from Finite Element Method (FEM) simulations. We present a novel message-passing GNN that encodes nodal strain and edge distances between FEM mesh cells, and aggregates to obtain embeddings and combines the decoded embeddings with the nodal strains to predict stress tensors on graph nodes. The GNN is trained on subgraphs generated from FEM mesh graphs, in which the mesh cells are converted to nodes and edges are created between adjacent cells. We apply the trained GNN to periodic polycrystals with complex geometries and learn the strain-stress maps based on crystal plasticity theory. The GNN is accurately trained on FEM graphs, in which the $R^2$ for both training and testing sets are larger than 0.99. The proposed GNN approach speeds up more than 150 times compared with FEM on stress predictions. We also apply the trained GNN to unseen simulations for validations and the GNN generalizes well with an overall $R^2$ of 0.992. The GNN accurately predicts the von Mises stress on polycrystals. The proposed model does not overfit and generalizes well beyond the training data, as the error distributions demonstrate. This work outlooks surrogating crystal plasticity simulations using graph data.

\vspace{25pt}


\end{abstract}


\end{frontmatter}

\clearpage
\tableofcontents

\clearpage


\nomenclature{\(\bf F\)}{Deformation gradient tensor}
\nomenclature{\(\mathbf{F}^e\)}{Elastic portion of deformation gradient tensor}
\nomenclature{\(\mathbf{F}^p\)}{Plastic portion of deformation gradient tensor}
\nomenclature{\(\tau\)}{Shear stress}
\nomenclature{\(\bf f\)}{Body force}
\nomenclature{\(\mathbf{v}^e\)}{Elastic stretch}
\nomenclature{\(\mathbb{C}\)}{Stiffness tensor}
\nomenclature{\(\epsilon^e\)}{Elastic strain}
\nomenclature{\(\epsilon^p\)}{Plastic strain}
\nomenclature{\(\mathbf{r}^*\)}{Lattice rotation}
\nomenclature{\(\bm{x}\)}{Current configuration}
\nomenclature{\(\mathbf{X}\)}{Reference configuration}
\nomenclature{\(\hat{\mathbf{L}}^p\)}{Plastic velocity gradient}
\nomenclature{\(\epsilon\)}{Lagrangian strain tensor}
\nomenclature{\(\hat{\mathbf{D}}^{p'}\)}{Plastic deformation rate}
\nomenclature{\(\hat{\mathbf{W}}^{p}\)}{Plastic spin}
\nomenclature{\(\hat{\mathbf{p}}^{\alpha}\)}{Symmetric part of the Schmid tensor for the $\alpha$ slip system}
\nomenclature{\(\hat{\mathbf{q}}^{\alpha}\)}{Skew part of the Schmid tensor for the $\alpha$ slip system}
\nomenclature{\(\dot{\gamma}^\alpha\)}{ Shearing rate of the $\alpha$ slip system}
\nomenclature{\(\dot{\gamma}_0\)}{Fixed-state strain rate scaling coefficient}
\nomenclature{\(h_0\)}{Strength hardening rate coefficient}
\nomenclature{\(\dot{g}^\alpha\)}{Strength of the $\alpha$ slip system}
\nomenclature{\(\mathbf{v}\)}{Velocity vector of a point in the current configuration}
\nomenclature{\(\bar{\mathbf{v}}\)}{Imposed velocity vector on the surface}
\nomenclature{\(\hat{\mathbf{s}}^\alpha\)}{Slip direction for the $\alpha$ slip system}
\nomenclature{\(\hat{\mathbf{m}}^\alpha\)}{Normal to the slip plane for the $\alpha$ slip system}
\nomenclature{\(g_s\)}{Saturation strength}
\nomenclature{\(g_0\)}{Initial slip system strength}
\nomenclature{\(n\)}{Nonlinear Voce hardening exponent}
\nomenclature{\(m\)}{Fixed-state strain rate sensitivity}
\nomenclature{\(\mathcal{G}\)}{Graph object}
\nomenclature{\(V\)}{Vertices of graph}
\nomenclature{\({V}_{\rm sub}\)}{Vertices (nodes) of subgraph}
\nomenclature{\(E\)}{Edges of graph}
\nomenclature{\(\bf p\)}{Material parameters}
\nomenclature{\(\mathcal{M}_{ij}\)}{Message information obtained on edges}
\nomenclature{\(\ell_{ij}\)}{Edge link length between mesh cells}
\nomenclature{\(\sf MSG\)}{Message function to pass message from nodes to edges}
\nomenclature{\(\sf MLP\)}{Multi-layer perceptron}
\nomenclature{\(\bigoplus\)}{Aggregation operator}
\nomenclature{\(\mathbf{h}_{ij}\)}{Node information on edge $i{-}j$}
\nomenclature{\(\mathbf{h}_{i}\)}{Node information on node $i$}
\nomenclature{\(\Xi\)}{Output decoded information on nodes}
\nomenclature{\(\mathcal{L}\)}{Loss function for the optimization problem}
\nomenclature{\(\phi\)}{MLP within the message-passing layer}
\nomenclature{\(\Phi\)}{MLP outside the message-passing layer on nodes}
\nomenclature{\(\mathcal{N}(i)\)}{Neighboring node list for node $i$}
\nomenclature{\(\sigma_{\rm vM}\)}{von Mises stress}
\nomenclature{\(\xi_\text{train}\)}{Subgraph ratio of the full graph}
\nomenclature{\(N_G\)}{Number of training graphs}
\nomenclature{\(\mathbb{N}\)}{Number of nodes in a graph}
\nomenclature{\(\mathbb{M}\)}{Number of edges in a graph}
\nomenclature{\(\mathbf{c}^{\tt ind}\)}{Connection indices of the subgraph}
\begin{spacing}{0.01}
\printnomenclature
\end{spacing}

\section{\label{intro}Introduction}
Plasticity refers to the permanent deformation of solid materials under external load, of which has been researched for more than 100 years. The earliest efforts include works of von Mises \cite{vonMises1913} and Huber \cite{Huber1904} to phenomenologically capture yield criteria. The {\em flow} process describes the post-yielding behavior, in which dislocation plays a significant role. Accurate predictions of plastic deformation are crucial for various practical applications, such as optimizing metal forming processes \cite{metal_froming}, designing materials with specific properties, e.g., fatigue resistance \cite{fatigue_resistance}), controlling semiconductor interconnects \cite{semiconductor_interconnects}, and controlling metal 3D printing processes \cite{additive_manufacturing}. These applications demand accurate and efficient digital twins characterizing crystal plasticity.

Due to decades of effort in understanding plasticity, constructing the constitutive model for polycrystals is still an active and ongoing research area due to (1) There are various ways to pose plasticity mechanisms characterizing the plastic deformation in continuum models such as temperature and rate dependence, anisotropy, etc.~\cite{plasticity_state_based, drucker_prager, state_rate_plas_model}; (2) By nature, plasticity is a multiscale problem, where numerous mechanisms contribute to the overall plastic behavior, such as single crystal dislocation \cite{Sills2018PRL}, inter-grain friction \cite{Luan2021_friction}, and grain boundary interactions \cite{Chen2024_grain_interaction_plas}, making it a challenging task to craft plasticity models integrate phenomena occurring at various scales; (3) The high computational expense associated with accurately simulating polycrystal plasticity using numerical methods such as finite elements \cite{fem_plas_time_1, Needleman1985_fem_plas_time_2, Kalidindi1992_plas_time_3}. The computational cost is mainly attributed to the path dependence and nonlinear nature of plasticity.

The recent developments of data-driven modeling for physical models could potentially task the high computational cost and surrogate plasticity models, considering their demonstrated success in fluid mechanics \cite{Raissi2020_science_fluid, Zhai2022_PINN_bubble}, heat transfer \cite{Guo2023_npj_phonon_pred_ML, Guo2023_npj_phonon_pred_ML_2}, and design optimization \cite{Zhai2022_BO_design_acs, Zhai2023_BO_design_jmbbm, Zhai2024_APLMat}. In the subgrain scale, mechanical responses can be predicted by combining machine learning and dislocation simulation data \cite{stress_strain_dd, ddd_sim_stress_strain_curve}. It has also been shown convolutional neural networks (CNN), graph neural networks (GNN), and general regression methods (e.g., Ridge, Lasso) can be applied to molecular dynamics (MD) simulations to predict mechanical responses and corresponding material properties \cite{md_cnn_yield_shear_mod, gnn_md_stress_ener, mat_prop_md_sim}. Based on grain representations, GNN has been used to predict the magnetic properties of polycrystalline graphs~\cite{gnn_polycry_magne}. Using experimental data, mechanical responses can be predicted from 2D images of static structures and CNN~\cite{cnn_struct_mech_image} or graph convolutional networks~\cite{gcn_mech_prop_exp}. Pagan et al.~\cite{Pagan2022} demonstrated that GNN can learn the anisotropic elastic response of alloys. In the continuum scale, CNN has been used to predict stress-strain responses~\cite{cnn_stress_strain_crystal_plas}. GNN has been used to learn mesh-based time-dependent PDEs~\cite{pde_mesh_gnn}. Notably, Mozaffar et al.~\cite{PNAS_plasticity_RNN} demonstrate that recurrent neural networks can learn path-dependent plasticity, and Fuhg et al.~\cite{Fuhg2022_ML_crystal_plasticity} show that partially input convex neural networks can predict plane stress macroscopic yield as a function of crystallographic texture.

This paper aims to demonstrate the capability of GNN in capturing the mechanical responses of polycrystals in the plastic regime, with two key innovations: (1) Handling complex geometry in polycrystal plasticity, and (2) Leveraging subgraph training for more efficient learning. Using open-source polycrystal generation and finite element method (FEM) software, {\em Neper} \& {\em FEPX} to generate meshes and conduct numerical simulations of periodic polycrystals~\cite{neper_cmame_ppr, fepx_arxiv}, the goal is to develop accurate and generalizable surrogate plasticity models based on finite element calculations~\cite{gnn_mesh_stress_strain, Gulakala2023}. Developing such surrogate models has three main challenges: (1) The generated finite element meshes have varying degrees of freedom (DoF) for different polycrystal geometries. Traditional regression tools, such as Gaussian processes or neural networks, typically rely on a fixed number of training points. (2) The spatial connectivity between finite element mesh cells preserves important physical features, i.e., physical properties are passed between adjacent mesh cells during the finite element calculations. Matrix-based data struggles to preserve such geometric relationships. (3) The data size is large; each 10-grain polycrystal mesh contains approximately 10,000 mesh cells, making the training a computationally expensive task.

To tackle the first problem, we propose using GNN to handle data with different DoFs. Since GNN can be trained on graphs with different numbers of nodes \& edges, meshes with different sizes can be potentially handled. This also helps us solve the second problem since GNN can handle the connectivity within the data. The connections between mesh cells preserve the spatial feature of the polycrystals. We generate graphs in which cells are converted to nodes where the adjacent cells are connected. To tackle the third problem, we propose training the GNN on subgraphs of finite element meshes. In this paper, we hope to combine the proposed approaches and explore stress predictions on polycrystals using GNN.

The paper is arranged as follows: In Section \ref{sec_crystal_plas} we present the formulation of the crystal plasticity model and the data generation process. In Section \ref{sec_gnn} we present the mathematical model and training details of the message-passing GNNs, explaining the details of subgraph sampling and training, with additional explanations of the mesh-graph data conversion process. In Section \ref{sec_results} we present the results of the predictions on training and testing sets, analysis of comparing GNN with FEM, and further deployment on unseen datasets as validation. We briefly conclude the paper in Section \ref{sec_conclusion}.



\section{Crystal plasticity\label{sec_crystal_plas}}

\subsection{Mechanistic model and problem formulation}

\begin{figure}[htbp]
    \centering
    \includegraphics[width=0.7\linewidth]{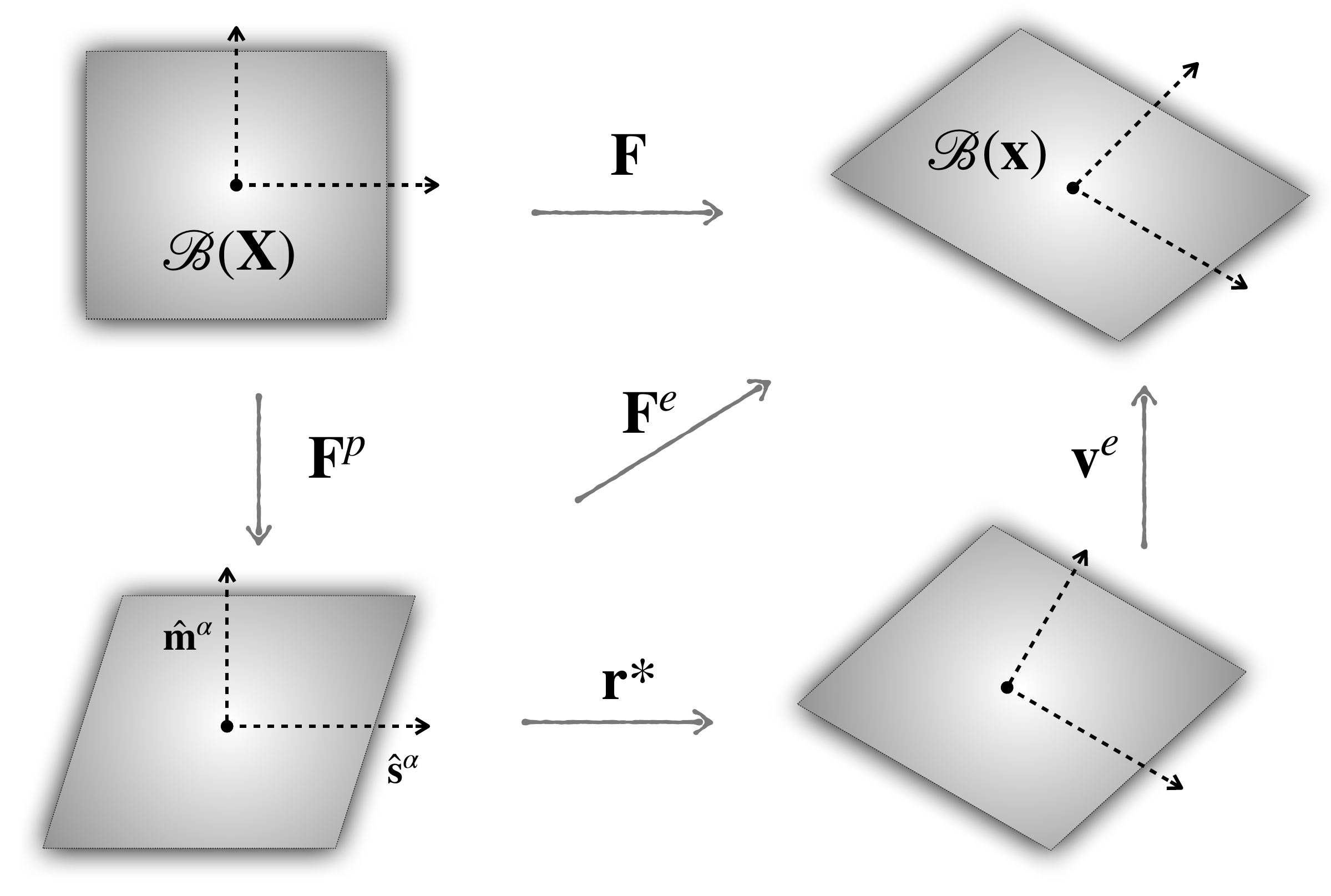}
    \caption{Schematic diagram for the decomposition in different configurations in crystal plasticity formulation. The visualization is inspired by Refs.~\cite{fepx_arxiv, huajian_crystal_plas_jmps}.}
    \label{plasticity schematic}
\end{figure}

Crystal plasticity models are employed \cite{phys_plas_def}, in which we use the general theory following Han et al. and the FEM implementation in {\em FEPX} \cite{huajian_crystal_plas_jmps, fepx_arxiv}. We begin with the deformation gradient tensor, defined as $\mathbf{F} = \frac{\partial \bm{x}}{\partial\mathbf{X}}$, can be decomposed into elastic and plastic parts:\begin{equation*}
    \mathbf{F} = \mathbf{F}^{e} \mathbf{F}^{p} = \mathbf{v}^e \mathbf{r}^* \mathbf{F}^p
\end{equation*}where the elastic gradient tensor can be decomposed to lattice rotation $\mathbf{r}^*$ and elastic stretch $\mathbf{v}^e$. $\mathbf{F}^p$ pertains plastic slip. $\bm{x}$ is the current configuration and $\bf X$ is the reference configuration. The general schematic of the theory is illustrated in Figure \ref{plasticity schematic}.

The polycrystal will generate a stress field distribution $\sigma$. Under loading, the local form of the equilibrium equation writes:
\begin{equation*}
    \nabla\cdot \mathbf{\sigma} + {\mathbf{f}} = 0
\end{equation*}where $\sigma$ is the Cauchy stress (or simply termed ``stress''). $\bf f$ is the body force vector, in our implementation $\mathbf{f} = 0$. The relationship between the Cauchy stress and the shear stress writes:\begin{equation*}
    \mathbf{\tau} = \left({\tt det}\left(\mathbf{v}^e\right) \right)\mathbf{\sigma}
\end{equation*}
For elastic deformations, the stress-strain relationship can be expressed as the generalized Hooke's law, which can be written as the \begin{equation}
    \sigma = \mathbb{C}\epsilon^e\label{eqn_linear_elas}
\end{equation}where $\mathbb{C} $ is the elastic moduli tensor (or stiffness tensor). $\mathbb{C} = \left[\mathcal{C}_{ij}\right]$ contains elastic constants to be specified in the simulation. 


After yield, the stress contributes to plastic flow, which can described by restricted slip. Here, $\hat{\mathbf{L}}^p$ is the plastic velocity gradient, which can be written in terms of the plastic slip:\begin{equation}
    \hat{\mathbf{L}}^p = \dot{\left(\mathbf{F}^p\right)} \left({\mathbf{F}^p}\right)^{-1}\label{eqn_plas_def_grad}
\end{equation} 

The Lagrangian strain tensor contains both the elastic and plastic contributions and can expressed in terms of elastic and plastic strain tensors:\begin{equation*}
    \begin{aligned}
        \mathbf{\epsilon}&=\frac{1}{2}\left({\mathbf{F}^e}^{\sf T}{\mathbf{F}^e} - \left(\mathbf{F}^p\right)^{\sf -T}\left({\mathbf{F}^p}\right)^{\sf -1}\right)\\
        & = \mathbf{\epsilon}^e + \mathbf{\epsilon}^p
    \end{aligned}
\end{equation*}

Using Schmid tensor's symmetric and skew part, the plastic deformation gradient in Eqn.~\eqref{eqn_plas_def_grad} can be written in terms of slip using Schmid tensor's symmetric and skew parts:\begin{equation*}
    \hat{\mathbf{L}}^p = \hat{\mathbf{D}}^{p'} + \hat{\mathbf{W}}^p
\end{equation*}where\begin{equation}
    \hat{\mathbf{D}}^{p'} = \sum_\alpha \dot{\gamma}^\alpha \hat{\mathbf{p}}^\alpha,\quad \text{and}\quad \hat{\mathbf{W}}^p = \dot{\left(\mathbf{r}^*\right)} \left(\mathbf{r}^*\right)^{\sf T} + \sum_\alpha \dot{\gamma}^\alpha \hat{\mathbf{q}}^\alpha\label{eqn sym and spin tensors}
\end{equation}

Here, $\hat{\mathbf{p}}^\alpha$ and $\hat{\mathbf{q}}^\alpha$ are defined as\begin{equation}
    \begin{aligned}
        \hat{\mathbf{p}}^\alpha = \hat{\mathbf{p}}^\alpha(\mathbf{q}) = \mathtt{sym}(\hat{\mathbf{s}}^\alpha \otimes \hat{\mathbf{m}}^\alpha)\\
        \hat{\mathbf{q}}^\alpha = \hat{\mathbf{q}}^\alpha(\mathbf{q}) = \mathtt{skw}(\hat{\mathbf{s}}^\alpha \otimes \hat{\mathbf{m}}^\alpha)
    \end{aligned}\label{eqn_schmid_skew_sym}
\end{equation}where $\hat{\mathbf{s}}^\alpha$ and $\hat{\mathbf{m}}^\alpha$ are the slip directions obtained after the kinetic decomposition visualized in Figure \ref{plasticity schematic}. Note that the symmetric and skew parts are expressed as: \begin{equation*}
    \begin{aligned}
        {\tt sym}(\cdot) = \frac{1}{2}\left[(\cdot) + (\cdot) \right]^{\sf T}\\
        {\tt skw}(\cdot) = \frac{1}{2}\left[(\cdot) - (\cdot) \right]^{\sf T}
    \end{aligned}
\end{equation*}
$\dot{\gamma}^\alpha$ is the slip system shearing rate. Here, the shearing rate relates to the resolved shear stress $\tau^\alpha$ via an assumed power law relationship:\begin{equation}
    \dot{\gamma}^\alpha = \dot{\gamma}_0 \left(\frac{|\tau^\alpha|}{g^\alpha}\right)^{\frac{1}{m}} \mathtt{sgn}(\tau^\alpha)\label{shear_stress_strain_eqn}
\end{equation}where $\dot{\gamma}_0$ is the fixed-rate strain rate scaling coefficient, $m$ is the rate sensitivity exponent. The resolved shear stress $\tau^\alpha$ is the projection of the crystal stress tensor onto the slip plane (in that particular slip direction) obtained via the Schmid tensor's symmetric part (Eqn.~\eqref{eqn_schmid_skew_sym}):\begin{equation*}
    \tau^\alpha = \mathtt{tr} \left(\hat{\mathbf{p}}^\alpha \tau'\right)
\end{equation*}

The evolution of slip system strength $g^\alpha$ can be characterized by hardening modulus $h_0$ and the initial strengths following a power law:\begin{equation*}
    \dot{g}^\alpha = h_0 \left(\frac{g_s(\dot{\gamma}) - g^\alpha}{g_s(\dot{\gamma}) - g_0}\right)^{n} \dot{\gamma}
\end{equation*}where $n$ is the nonlinear Voce hardening exponent. $g_s(\dot{\gamma})$ is the initial slip system saturation strength. $g_0$ is the initial slip system strength. $\dot{\gamma}$ is calculated as the summation of the slip shearing rates, related to resolved shear stresses (Eqn.~\eqref{shear_stress_strain_eqn}):\begin{equation*}
    \dot{\gamma} = \sum_\alpha \left|\dot{\gamma}^\alpha\right|
\end{equation*}

For the FEM implementations of this method, some numerical details are summarized in \ref{appendix fem alg}.

The boundary conditions (B.C.s) can be specified via \begin{equation*}
    \mathbf{v}(\mathbf{x}) = \bar{\mathbf{v}}
\end{equation*}as the velocity B.C.s. In our implementation in {\em FEPX} \cite{fepx_arxiv}, we apply fixed strain rate in $x$-direction, $\dot{\epsilon}_{xx} = 10^{-3}\rm\ s^{-1}$, which only acts on the $v_x$ components. The applied strain rates in other directions are all set to be zero. 

\subsection{Material parameters \& data generation\label{sec_data_gen_fem}}

The rate sensitivity exponent in Eqn.~\eqref{shear_stress_strain_eqn} is set to be $m = 0.02$. We employ an isotropic hardening type. The fixed-rate strain rate is $\dot{\gamma}_0 = 1$. The simulation targets a total strain of $\epsilon_{xx} = 0.01$, where the strain increment per step is 0.001. We generate 90 10-grain periodic polycrystals, in which the mesh is generated via {\em Neper} \cite{neper_cmame_ppr}. We used body-centered cubic (BCC) crystals with elastic constants $\mathcal{C}_{11} = 236.9$ [GPa], $\mathcal{C}_{12} = 140.6$ [GPa], and $\mathcal{C}_{44} = 116.0$ [GPa]. The hardening modulus $h_0 = 391.90$ [MPa] and the slip strengths are $g_0 =200$ \& $g_s = 335$ [MPa], respectively. The nonlinear Voce hardening exponent is $n=1$. The 90 simulation results are then converted to graphs, of which 80\% (72 graphs) are selected for the training, and the remaining (18 graphs) are considered as the testing sets. See Refs. \cite{fepx_arxiv, fepx_original} for details and related FEM implementation.

\begin{figure}[htbp]
    \centering
    \includegraphics[width=0.65\linewidth]{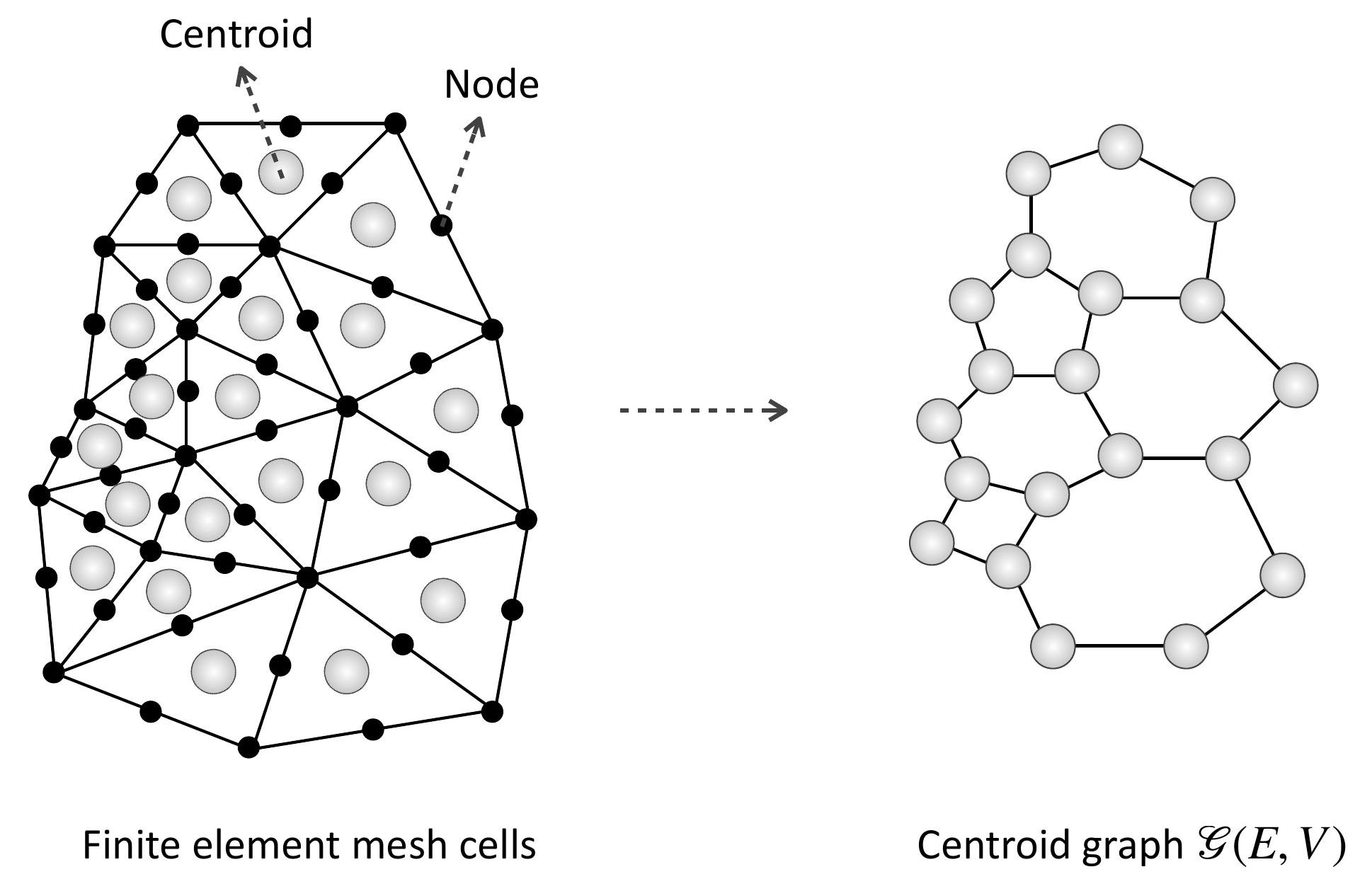}
    \caption{Schematic illustration of the procedures for converting FEM meshes to graphs. FEM element cell centroids are treated as nodes, and neighboring cells (sharing 3 common nodes for {\tt `tetra10'} elements) share an edge. }
    \label{fig_mesh_graph}
\end{figure}

In our formulation, we hypothesize that the finite element meshes can be formulated as a graph $\mathcal{G} = \mathcal{G}(V, E)$. $V\in\mathbb{R}^\mathbb{N}$ \& $E\in\mathbb{R}^\mathbb{M}$ are vertices and edges of the graph\footnote{we are using the terms {\em vertex} and {\em node} interchangeably in this manuscript.}, where $V = V\left(\epsilon_i \mapsto \sigma_i\right)$ are the node features (on the finite element mesh node $i$) and $E = E\left(\ell_{ij}\right)$ (Euclidean distances of mesh cells, on edge $i$-$j$ that connects nodes $i$ \& $j$). $\mathbb{M}$ \& $\mathbb{N}$ are the number of edges and nodes. Each cell of the finite element mesh is considered a node. The edges are constructed according to the connectivity of the nodes.
To enhance training efficiency, the strain data is rescaled by multiplying $10^4$, and the Euclidean norms are rescaled by multiplying $10^3$, making all the feed in-out data have a similar scale with the stress data ($\sim10^3$). Figure \ref{fig_mesh_graph} indicates how the finite element meshes are converted to graph data for training. For the created {\tt tetra10} mesh for polycrystals, the centroids were converted to graph nodes (or vertices). For two adjacent mesh cells sharing 3 common nodes (or one mesh edge), an edge is being created on the graph\footnote{Note that this also respects the conformality in finite element mesh.}. Similar approaches are also employed for multiscale plasticity and topology optimization \cite{VlassisSun2023_mesh_graph, Gavris_EML_2024}. Some discussions on this graph conversion method are provided in \ref{appendix graph conversion}.

Here, we aim to learn the nodal map from total strain to stress, i.e., $\mathfrak{M}: \mathbf{\epsilon}\in\mathbb{R}^6 \mapsto \sigma\in\mathbb{R}^6$. We want to use the GNN to surrogate the model $\mathfrak{M}$. By converting the mesh to graphs, one can construct mapping for polycrystals of irregular complex geometries since the learning is independent of the dimensions of the data. The overall strain-stress map for the physics-based model can be simplified in a form:\begin{equation}
    \sigma^{FEM}_i \equiv \sigma_i(\mathbf{x}) = \mathfrak{M}\left([\mathbf{\epsilon}_i(\mathbf{x}), \mathbf{F}]; \mathbf{p}\right)\label{eqn_fem_stress}
\end{equation}where $\mathbf{p} = (\mathbb{C}, g_0, g_s, h_0, m, n, ...)$ subsumes all the related material parameters used in the simulation. The model $\mathfrak{M}(\cdot)$ takes strain $\epsilon_i$ and the configurational map $\bf F$, and $\bf p$ as input and predict stress $\sigma_i$ according to the equations presented above.

\begin{figure}[htbp]
    \centering
    \includegraphics[width=.9\linewidth]{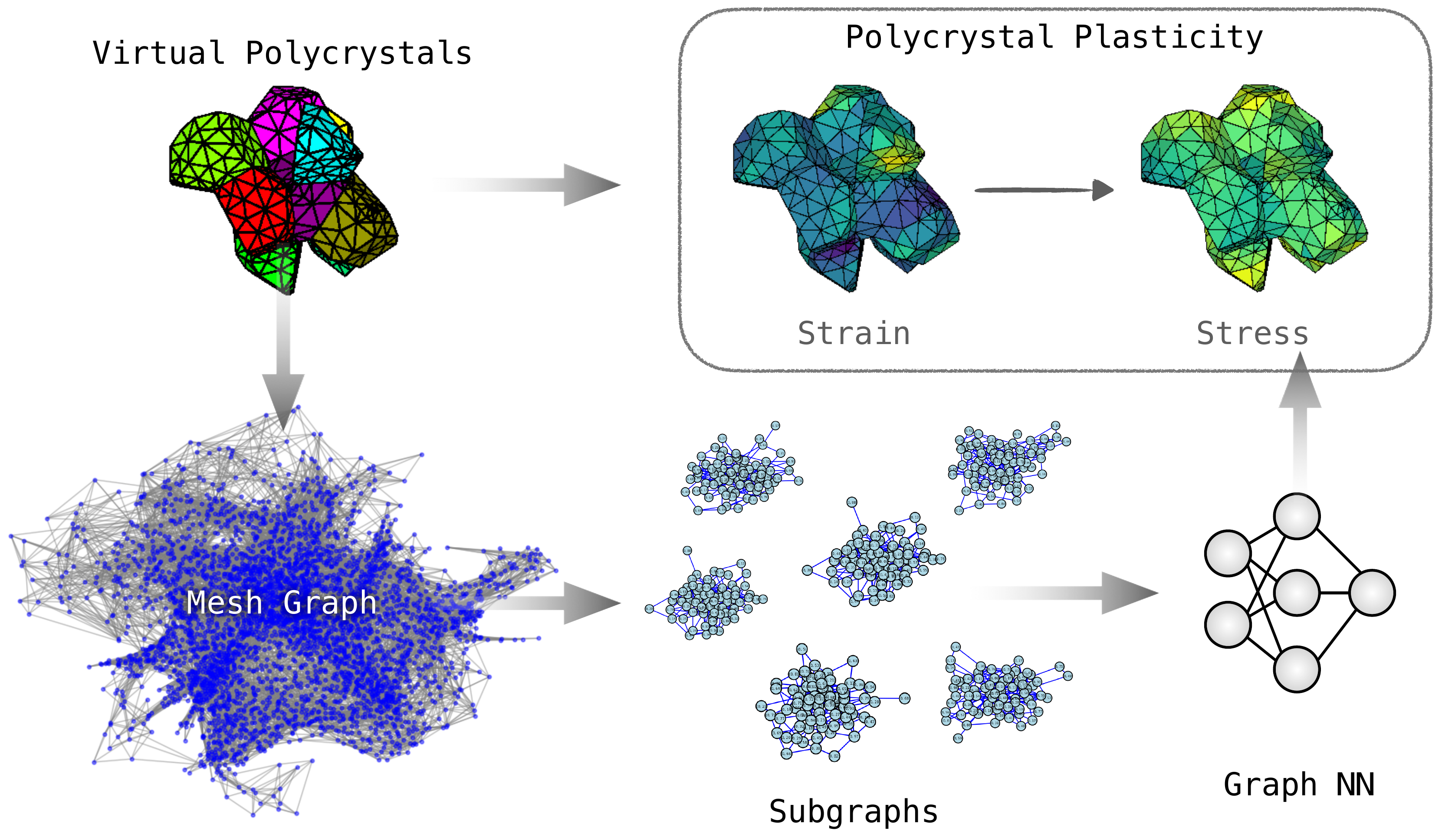}
    \caption{General schematic of the workflow for using GNN to learn polycrystal plasticity. Virtual polycrystals are generated using {\em FEPX}. It is converted to mesh graphs based on finite element cells. The subgraphs are extracted to train the GNN. The GNN is then deployed to surrogate polycrystal plasticity simulations.}
    \label{fig 1 schematic}
\end{figure}

In Eqn.~\eqref{eqn_fem_stress}, $\sigma_i$ contains six stress elements are defined as $\left\{\sigma_1, \sigma_2, \sigma_3, \sigma_4, \sigma_5, \sigma_6\right\}\equiv\left\{\sigma_{11}, \sigma_{12}, \sigma_{13}, \sigma_{22}, \sigma_{23}, \sigma_{33}\right\}\in\mathbb{R}^6$ (same for the strain components) for the overall stress \& strain tensor elements in the FEM implementation. Note that $\{x,y,z\}$ and $\{1,2,3\}$ are used interchangeably in the subscripts.

\section{Message-passing graph neural networks\label{sec_gnn}}

\subsection{Message-passing on edges for nodal inference\label{sec_gnn_basics}}
Message-passing GNN learns the data relationship on graphs by passing the message from edges to nodes (vertex) and conducting nonlinear regression (using multi-layer perceptron, MLP) in the feature space. One begins with preparing the ``messages'' on edges, where the accumulated nodal message $\Tilde{\mathcal{M}}^\epsilon_{ij}$ (on edge) and edge message $\Tilde{\mathcal{M}}^\ell_{ij}$ for edge $i{-}j$ can be written as:\begin{equation}
    \begin{aligned}
        \Tilde{\mathcal{M}}^\epsilon_{ij} = {\sf MSG}^{(n)}(\{\mathbf{\epsilon}_{i}^{in},\ \mathbf{\epsilon}_{j}^{in}\}),\quad
        \Tilde{\mathcal{M}}^\ell_{ij} = {\sf MSG}^{(e)}(\{\mathbf{\ell}_{ij}\})
    \end{aligned}\label{mess_lay_eqn}
\end{equation}where $\Tilde{\mathcal{M}}^\epsilon_{ij}\in \mathbb{R}^{\mathbb{M}\times6}$ (passing the information of strains $\epsilon$) and $\Tilde{\mathcal{M}}^\ell_{ij}\in \mathbb{R}^{\mathbb{M}\times1}$ (information of Euclidean distances $\ell$). Here, $6$ and $1$ are the feature space dimensions for nodes and edges. $\mathbf{\epsilon}_i^{in}$ and $\mathbf{\epsilon}_j^{in}$ are the nodal strains (input property) for nodes $i$ \& $j$; and $\ell_{ij}$ are the edge input property, i.e., the mesh link length of edge $i{-}j$.

The nodal end edge messages are passed to the embedding dimension via two separate MLPs and output $\mathbf{h}_{ij}^{(n)}$ and $\mathbf{h}_{ij}^{(e)}$ with dimension $\mathbb{R}^{\mathbb{M}\times {\sf emb}}$. $\bigoplus_j$ is the aggregation operator that sums over the neighboring node and edge information for node $i$. The predicted embedding output is then concatenated into $\Tilde{\mathbf{h}}_{ij}$, with dimension $\mathbb{R}^{\mathbb{M}\times(2{\sf emb})}$:\begin{equation}
    \begin{aligned}
        \mathbf{h}_{ij}^{(n)} = {\sf MLP}^{(\textsc{n-Enc})}(\Tilde{\mathcal{M}}^\epsilon_{ij}),&\quad \mathbf{h}_{ij}^{(e)} = {\sf MLP}^{(\textsc{e-Enc})}(\Tilde{\mathcal{M}}^\ell_{ij}),\\
        \Tilde{\mathbf{h}}_{i}^{(n)} = \bigoplus_{j\in\mathcal{N}(i)} \left(\left\{\mathbf{h}_{ij}^{(n)} \right\}\right),&\quad \Tilde{\mathbf{h}}_{i}^{(e)} = \bigoplus_{j\in\mathcal{N}(i)} \left(\left\{\mathbf{h}_{ij}^{(e)} \right\}\right)
    \end{aligned}\label{eqn_mess_pass}
\end{equation}where $\mathcal{N}(i)$ stands for the neighboring node list for node $i$. To pass the prediction on nodes, aggregation is being conducted for the decoded information in the message-passing layer. The output is then concatenated and fed into the decoding MLP:\begin{equation*}
    \begin{aligned}
        {\Xi}_{i} ={\sf MLP}^{(\textsc{Dec})}\left(\left\{\Tilde{\mathbf{h}}_{i}^{(n)},\ \Tilde{\mathbf{h}}_{i}^{(e)} \right\}\right)
    \end{aligned}\label{aggregation_eqn}
\end{equation*}This process is the {\em message-passing} from edges to nodes. This aggregation is being done in the embedding dimension and the output $\Xi_i\in\mathbb{R}^{\mathbb{N}\times1}$ is the properties on the node. The prediction follows an ``equation-MLP'' using the given nodal information and decoded edge information (on the node):\begin{equation}
    \begin{aligned}
        \Tilde{\mathbf{\sigma}}_i = {\sf MLP}^{(\textsc{Eqn})}\left(\left\{ \mathbf{\epsilon}^{in}_i, \Xi_{i}\right\}\right)
    \end{aligned}\label{pred_lay_eqn}
\end{equation}where ${\Tilde{\mathbf{\sigma}}}_i\in\mathbb{R}^{\mathbb{N}\times6}$ are the final stress predictions for the supervised learning target, which is the optimization goal $\Tilde{\mathbf{\sigma}}_i\sim\mathbf{\sigma}_i$.

The training uses mean-squared error (MSE) as the objective $\mathcal{L}$ parameterized by trainable variables $\Theta$ for supervised training. ${\tt dim}()$ denotes the dimension of the given data. The optimization problem writes:\begin{equation}
    \begin{aligned}
        {\rm arg}\min_{\Theta}\mathcal{L}(\Theta),\\
        \mathcal{L} = \frac{1}{ {\tt dim}(\mathbf{\sigma})}\sum_{i\in {\tt dim}(\mathbf{\sigma})}{\left(\Tilde{\mathbf{\sigma}}_i - \mathbf{\sigma}_i\right)^2}
    \end{aligned}\label{opt_eqn}
\end{equation}where MSE is being calculated on all the nodes on the graph, ${\tt dim}(\sigma)=\mathbb{N}$. For a graph, ${\tt dim}()$ indicates the number of nodes.

To summarize, from Eqns.~\eqref{mess_lay_eqn}$\sim$\eqref{pred_lay_eqn}, the overall model can be simplified as a surrogate model for the $\epsilon\mapsto\sigma$ map:\begin{equation}
\mathbf{\sigma}^{GNN}_i\equiv\Tilde{{\mathbf{\sigma}}}_i = \Phi \left( \mathbf{\epsilon}_i^{in}, \bigoplus_{j \in \mathcal{N}(i)} \, \phi\left(\left\{\mathbf{\epsilon}_i^{in}, \mathbf{\epsilon}_j^{in}\right\},\mathbf{\ell}_{ij}\right) \right)\label{eqn_gnn_stress}
\end{equation}where $\Phi$ and $\phi$ represent (combination of) different MLPs. This formula follows the generalized formula for message-passing GNN. The prediction is estimated based on the comparison of stresses predicted by GNN and FEM denoted in Eqns.~\eqref{eqn_fem_stress} and \eqref{eqn_gnn_stress}. The performance of the model is evaluated using the coefficient of determination (denoted as \( R^2 \)), which quantifies how well the predicted stress values from the GNN model match the true FEM values. The \(R^2\) score is calculated as: 
\[R^2 = 1 - \frac{\sum_{i=1}^{\mathbb{N}} (\sigma_i^{FEM} - \sigma_i^{GNN})^2}{\sum_{i=1}^{\mathbb{N}} (\sigma_i^{FEM} - \overline{\sigma}^{FEM})^2}\]
where \( \sigma_i^{FEM} \) are the stress values predicted by the FEM model, \( \sigma_i^{GNN} \) are the stress values predicted by the GNN model, \( \overline{\sigma}^{FEM} \) is the mean value of the true FEM stress values, and \(\mathbb{N}\) is the number of data points. The \(R^2\) value typically ranges from 0 to 1, with a value of 1 indicating perfect agreement between the predicted and true values, and a value closer to 0 indicating poor prediction performance.

\subsection{Model framework and training algorithm\label{sec_gnn_model}}

\begin{figure}[htbp]
    \centering
    \includegraphics[width=.85\linewidth]{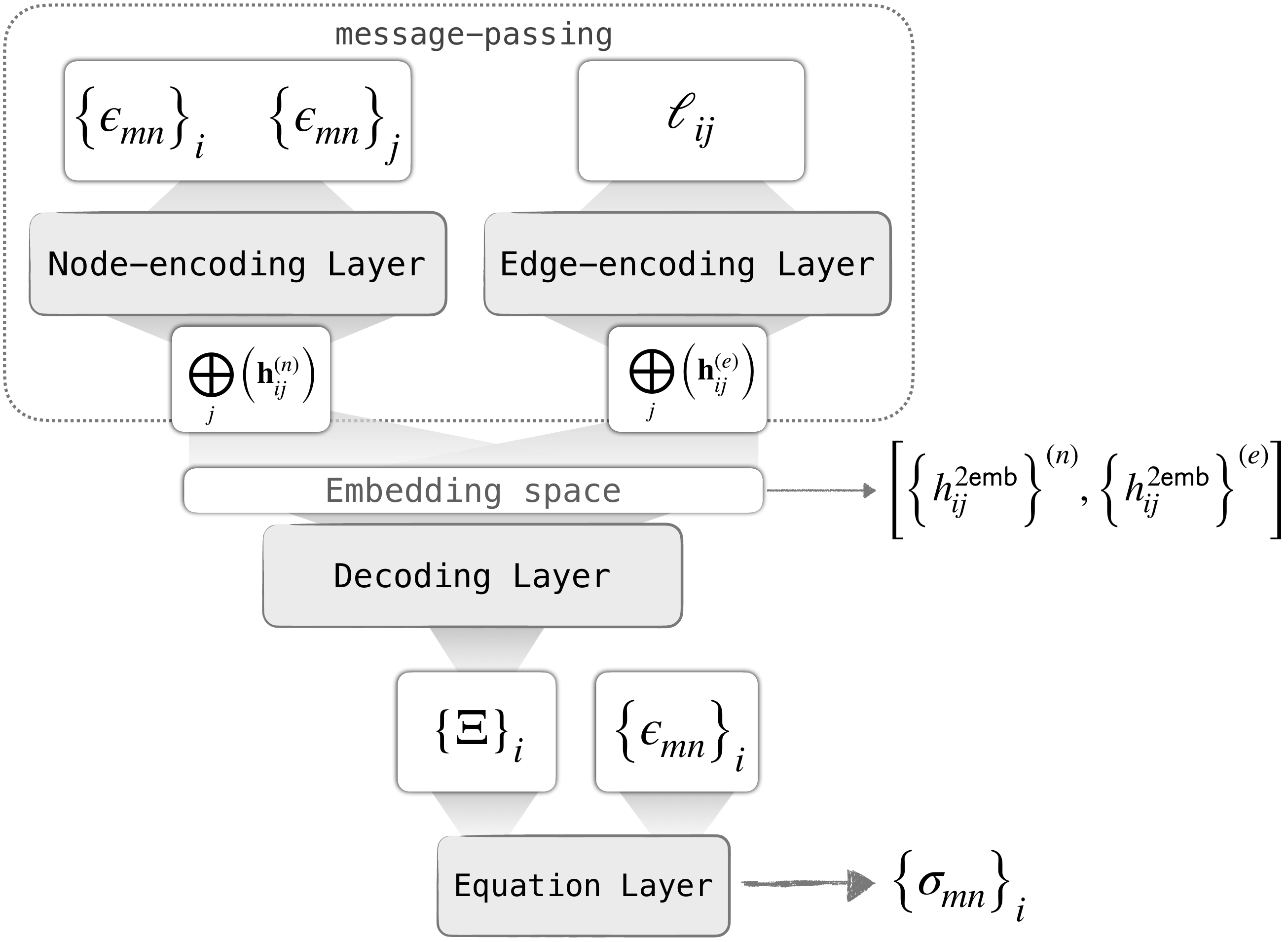}
    \caption{The general architecture for the GNN. The node-encoding layer takes the strains on neighboring nodes for the input edges, and the edge-encoding layer takes the mesh cell link length (i.e. Euclidean norm of mesh cells). The combined outputs are then fed input the embedding space ($\mathbb{R}^{2{\sf emb}}$). $\left\{h_{ij}^{2 \sf emb}\right\}^{(n)}$ and $\left\{h_{ij}^{2 \sf emb} \right\}^{(e)}$ are the decoded nodal and edge information in the embedding space. The output data is then fed input to the decoding layer that maps $\mathbb{R}^{2{\sf emb}}$ to $\mathbb{R}^4$. The output of the decoding layer is then passed to the message-passing operator (i.e. $\bigoplus$), where the decoded messages are passed on nodes. The edge information on nodes $\left\{\epsilon_{mn}\right\}_i$ are then combined to put into the equation layer, to predict the corresponding stress components $\left\{\sigma_{mn}\right\}_i$.}
    \label{fig 2 gnn}
\end{figure}

Figure \ref{fig 2 gnn} illustrates the general architecture of our message-passing GNN. The node and edge-encoding layer takes in the nodal and edge properties on edge $i{-}j$, and output $\mathbf{h}^{(n)}_{ij}$ and $\mathbf{h}^{(e)}_{ij}$, which are then being operated by the aggregation operator to pass the properties from edges to nodes to obtain $ \Tilde{\mathbf{h}}_{i}^{(n)}$ and $ \Tilde{\mathbf{h}}_{i}^{(e)}$ in the embedding dimension (Eqn.~\eqref{eqn_mess_pass}). The concatenated output $\left\{\Tilde{\mathbf{h}}_{i}^{(n)}, \Tilde{\mathbf{h}}_{i}^{(e)}\right\}$ ($\in\mathbb{R}^{\mathbb{N}\times\sf2emb}$) is then sent to the decoding layer to $\Xi_i\in\mathbb{R}^{\mathbb{N}\times1}$. ${\Xi}_i$ and node input property $\mathbf{\epsilon}_i$ are then being concatenated and input to the equation layer to give the prediction that aims to approximate $\mathbf{\sigma}_i$ (Eqn.~\eqref{pred_lay_eqn}). The subscripts $()_{mn}$ denote the elements in the stress \& strain tensors, and $()_{ij}$ denotes the connection of nodes in graphs.

There is 1 hidden layer for the {\em message-passing MLP} ($\phi$) and {\em equation MLP} ($\Phi$) respectively. The hidden dimension of the MLP ${\sf emb}=31$. As noted, the input and output dimensions are 6, that is, $\epsilon_i\mapsto\sigma_i, i=1,2,\cdots,6$. {\tt ReLU} activation function is used for $\Phi$ and {\tt tanh} activation function is used for $\phi$. Note that data is ``activated'' twice in $\phi$ for node- and edge-encoding layers respectively. Since there is only one hidden layer, the simple and light GNN model also effectively prevents the issue of oversmoothing~\cite{gnn_oversmooth}. More details of the model can be seen in~\ref{appendix_gnn}.

\subsubsection{Subgraph sampling \& training}

We propose training GNN on subgraphs to learn the mapping. Let $\mathcal{G}_{\sf sub}=\mathcal{G}_{\sf sub}(V_{\sf sub}, E_{\sf sub})$ denote the subgraph extracted from the full graph $\mathcal{G}$ with randomly selected nodes, where $V_{\sf sub}\subseteq V$ \& $E_{\sf sub}\subseteq E$. Within $V_{\sf sub}$, let $\hat{V}_{\sf sub}$ be all the subgraph nodes that preserve full edges compared with $\mathcal{G}$. $\hat{V}_{\sf sub} \backslash {V}_{\sf sub}$ are the nodes that lose edges during the subgraph extraction process. Each finite element mesh graph contains $\sim10^5$ nodes in our implementation. For effective and efficient training of GNN, we propose training GNN on the subgraph, in which only the $\hat{V}_{\sf sub}$ and the connected edges are considered in the loss calculation, $\mathcal{L} = {\tt MSE}\left(\mathcal{G}_{\sf sub}\left(\hat{V}_{\sf sub}\right)\right)$. $\hat{V}_{\sf sub}$ can be selected by comparing the number of edges per node of $\mathcal{G}_{\sf sub}$ and $\mathcal{G}$ (based on the global node index). The filtered node indices (termed as {\em connection indices}) are denoted as $\mathbf{c}^{\tt ind}$. One then uses $\mathbf{c}^{\tt ind}$ to select ``{\em active nodes}'' to train based on the loss function.

\begin{figure}[htbp]
    \centering
    \includegraphics[width=0.7\linewidth]{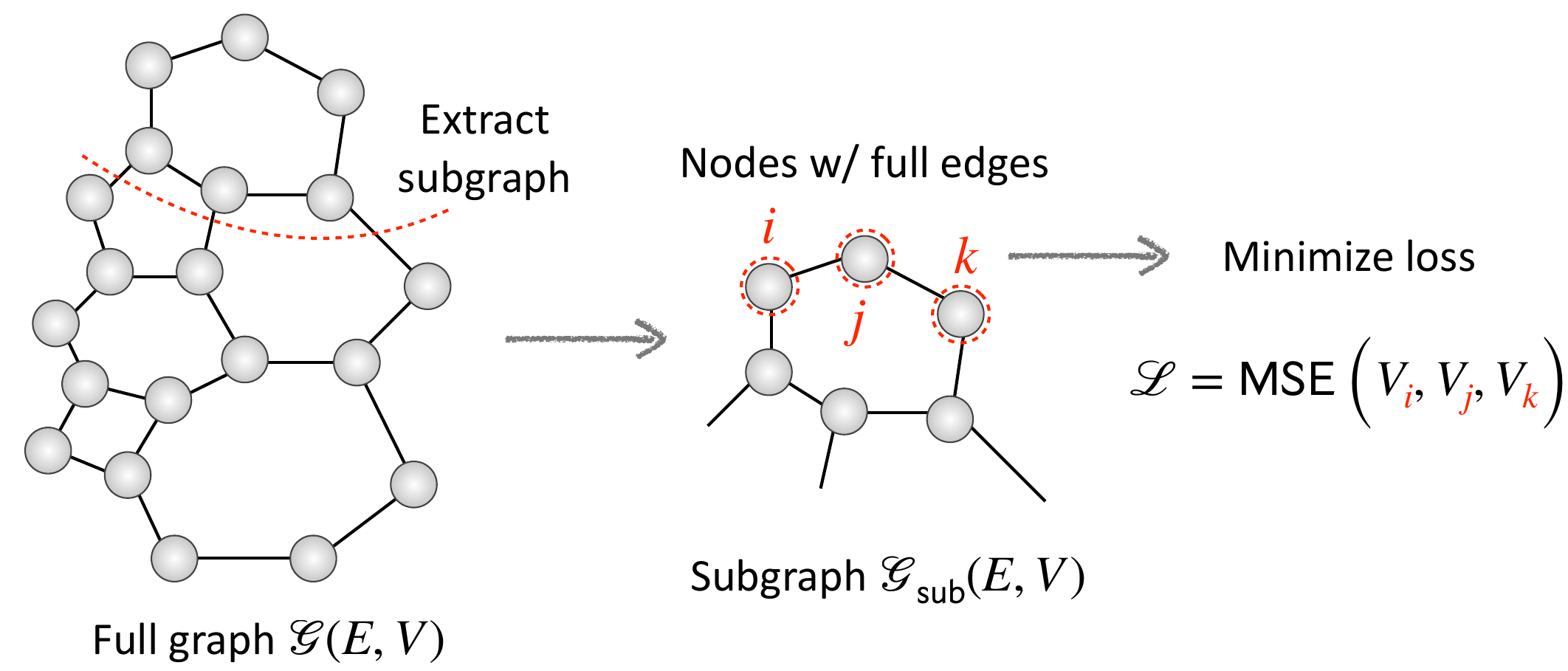}
    \caption{Schematic illustration for the proposed subgraph training method using the subgraph extracted from the full graph. The subgraph $\mathcal{G}_{\sf sub}$ are extracted from the sampled nodes from the full graph $\mathcal{G}$, in which the nodes containing full-edge information (e.g., $i, j, \&\ k$) were considered in the loss function during training.}
    \label{fig_sub_graph}
\end{figure}

\begin{figure}[htbp]
    \centering
    \includegraphics[width=0.75\linewidth]{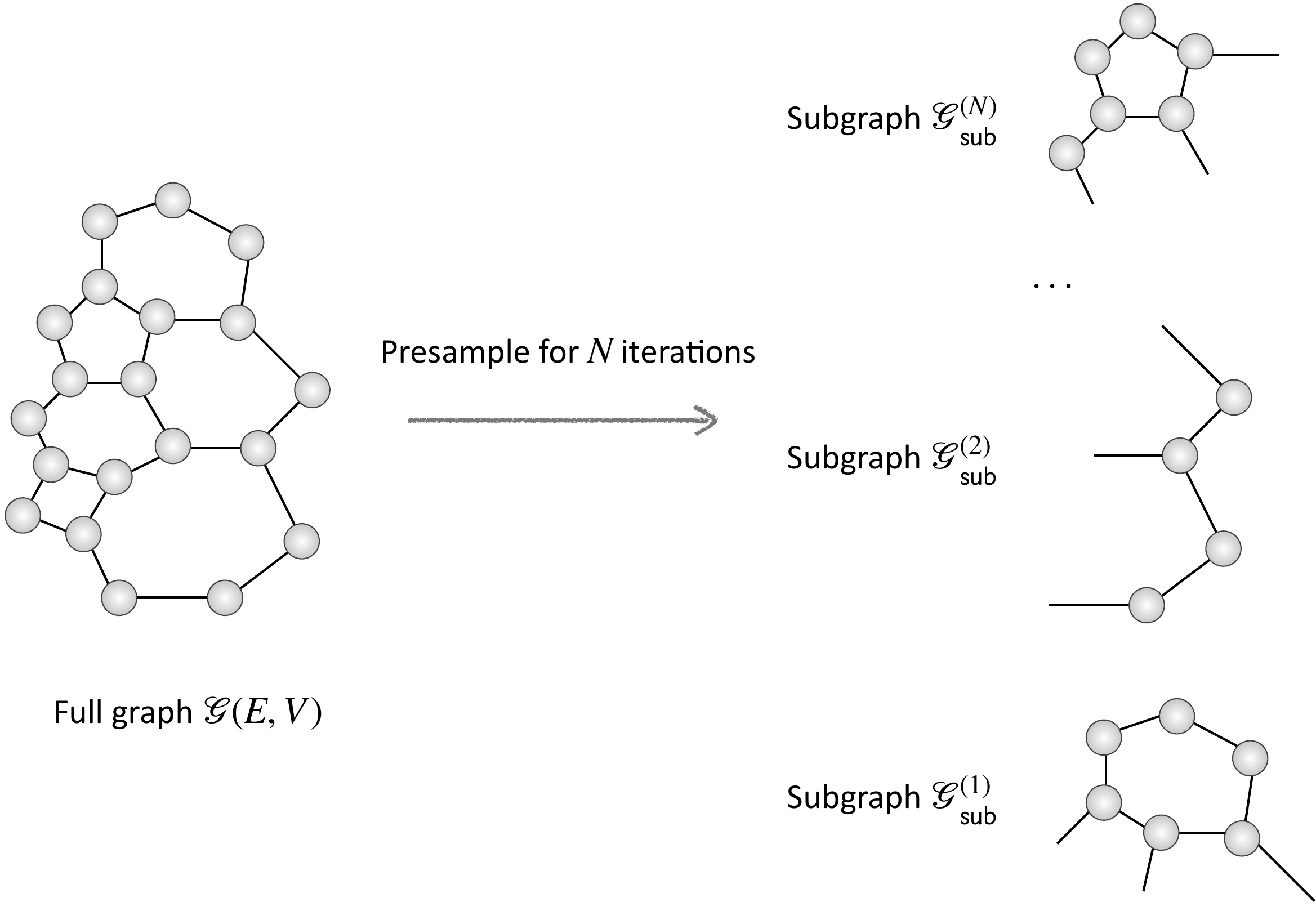}
    \caption{Schematic illustration of the presampling methods for subgraphs training to improve training efficiency and prediction accuracy. Subgraphs $\mathcal{G}_{\sf sub}^{N}$ are presampled for $N$ epochs for the full graph $\mathcal{G}$. All the FEM graphs in the training sets are presampled for $N$ epochs before training.}
    \label{fig_sub_graph_presample}
\end{figure}

Figure \ref{fig_sub_graph} illustrates the details of the subgraph training method to capture mapping on nodes. Based on a full graph converted from FEM meshes (Figure \ref{fig_mesh_graph}), one first extracts the subgraph from full graph\footnote{For details one could refer to \href{https://pytorch-geometric.readthedocs.io/en/2.4.0/_modules/torch_geometric/utils/subgraph.html}{\tt torch\_geometric.utils.subgraph}}, and then filter out the active nodes according to the connection index $\mathbf{c}^{\tt ind}$, exampled as $i$, $j$, $k$ in Figure \ref{fig_sub_graph} to train according to the loss function. To help the GNN to be more comprehensively trained according to this method, we propose presample the subgraphs before training, as shown in Figure \ref{fig_sub_graph_presample}. For a given full graph $\mathcal{G}$ to be trained on $N$ epochs, one sample of all the subgraphs in the training sets for the $i$-th epoch as $\mathcal{G}^{(i)}_{\sf sub}$. Essentially, for a full graph $\mathcal{G}$ being trained, the GNN is ``seeing'' new subgraphs for each epoch, in which it preserves the local feature (i.e., the constitutive map from strain to stress in our case) of the full graph.

\subsubsection{Training algorithm}

The training algorithm is shown in Algorithm~\ref{alg_gnn_train}. Finite element mesh-based graphs are stored in the form of {\tt Torch Geometric} tensors, containing nodal ($\epsilon$ \& $\sigma$) and edge properties ($\ell$). Subgrahs are extracted based on the training ratio $\xi_\text{train}$, specifying the ratio of the number of nodes selected from the full graph $\mathcal{G}$. We use $\xi_\text{train}=0.5$ for our training\footnote{half of the graph nodes are sampled from the full graph}. We pre-sample a set of subgraphs in the training set for each epoch and prepare a list of sampled subgraphs $\mathcal{B}_G$ for training implementation. Under each epoch, the unique subgraph sets per that epoch will be selected, in which the active nodes are selected based on $\mathbf{c}^{\tt ind}$ and fed into the loss function (Eqn. \eqref{opt_eqn}). Adam optimizer is selected for gradient-based optimization. The model is being trained on the 72 graphs for 1000 epochs. Accompanying our subgraph sampling and training method, we use a double loop structure to first loop the subgraph batch sampled per epoch and then loop over each subgraph to learn the local feature map (strain to stress) on the subgraphs. This hierarchical training enables the GNN to learn on the subgraph generated from different polycrystals at each model evaluation step, i.e., iteration.


\begin{algorithm}[htbp]
\caption{Training algorithms for message-passing GNN}\label{alg_gnn_train}
\begin{algorithmic}
\Require Graph data files containing $\mathcal{G} ({V}, {E})$ converted from finite element meshes, stored in the form of {\tt Torch Geometric} tensors; mapping the nodal input to outputs $\epsilon \in \mathbb{R}^{\mathbb{N}\times6} \oplus \ell\in \mathbb{R}^{\mathbb{M}\times1}\mapsto \sigma \in \mathbb{R}^{\mathbb{N}\times6}$. Number of training graphs $N_G$ ($=72$).
\State\hspace{-11pt}{\bf Hyperparameters:} Subgraph ratio: $\xi_\text{train}$; The embedding dimension $\sf emb$; Number of epochs ${\tt Epochs}$; Select optimizer $\tt Adam(\cdot)$; pre-sampled subgraphs list $\mathcal{B}_G({\tt Epochs})$ from the training graphs.
\State Load pretrained GNN model ${\tt GNN}[\cdot]$. \Comment{{\small\color{gray}optional based on existence}}
\For{$ep < {\tt Epochs}$}
\State $\mathcal{B}_G^{(ep)}\leftarrow \mathcal{B}_G(ep)$
\For{${\sf ID}_G$ in $N_G$}
\State $\mathcal{G}_{\sf sub} \leftarrow \mathcal{B}_G^{(ep)}\left(\mathsf{ID}_G\right) $\Comment{{\small\color{gray}obtain pre-sampled subgraph}}
\State ${\bf c}^{\tt ind}\leftarrow \mathcal{F}\left(\mathcal{G}_{\sf sub},\ \mathcal{G}\right)$.\Comment{{\small\color{gray}$\mathcal{F}(\cdot)$: filtering function to sort connection indices}}

\State $\tilde{\sigma}\gets {\tt GNN}\left[\mathcal{G}_{\sf sub}\left(\epsilon, \ell\right)\right]$ \Comment{{\small\color{gray}based on defined GNN in Sec. \ref{sec_gnn_model}.}}
\State $\mathcal{L} \leftarrow {\tt MSE}(\Tilde{\sigma}[{\bf c}^{\tt ind}], \mathcal{\sigma}[{\bf c}^{\tt ind}])$\Comment{{\small\color{gray}only active nodes are included in the loss.}}
\State ${\tt GNN}(\Theta)\xleftarrow{\rm backward}\mathcal{L}$.\Comment{{\small\color{gray}backpropagation}}
\State Clips gradient norm, \& optimization: ${\rm arg}\min_{\Theta}\mathcal{L}$.\Comment{{\small\color{gray}\tt Adam($\cdot$)}}
\EndFor\State $ep {+=} 1$
\EndFor
\State Save the trained GNN model ${\tt GNN}$. \Comment{{\small\color{gray}requires the specified device for testing.}}
\end{algorithmic}
\end{algorithm}






\section{Results and discussions\label{sec_results}}

\subsection{Training and testing results}

\begin{figure}[htbp]
    \centering
    \includegraphics[width=.9\linewidth]{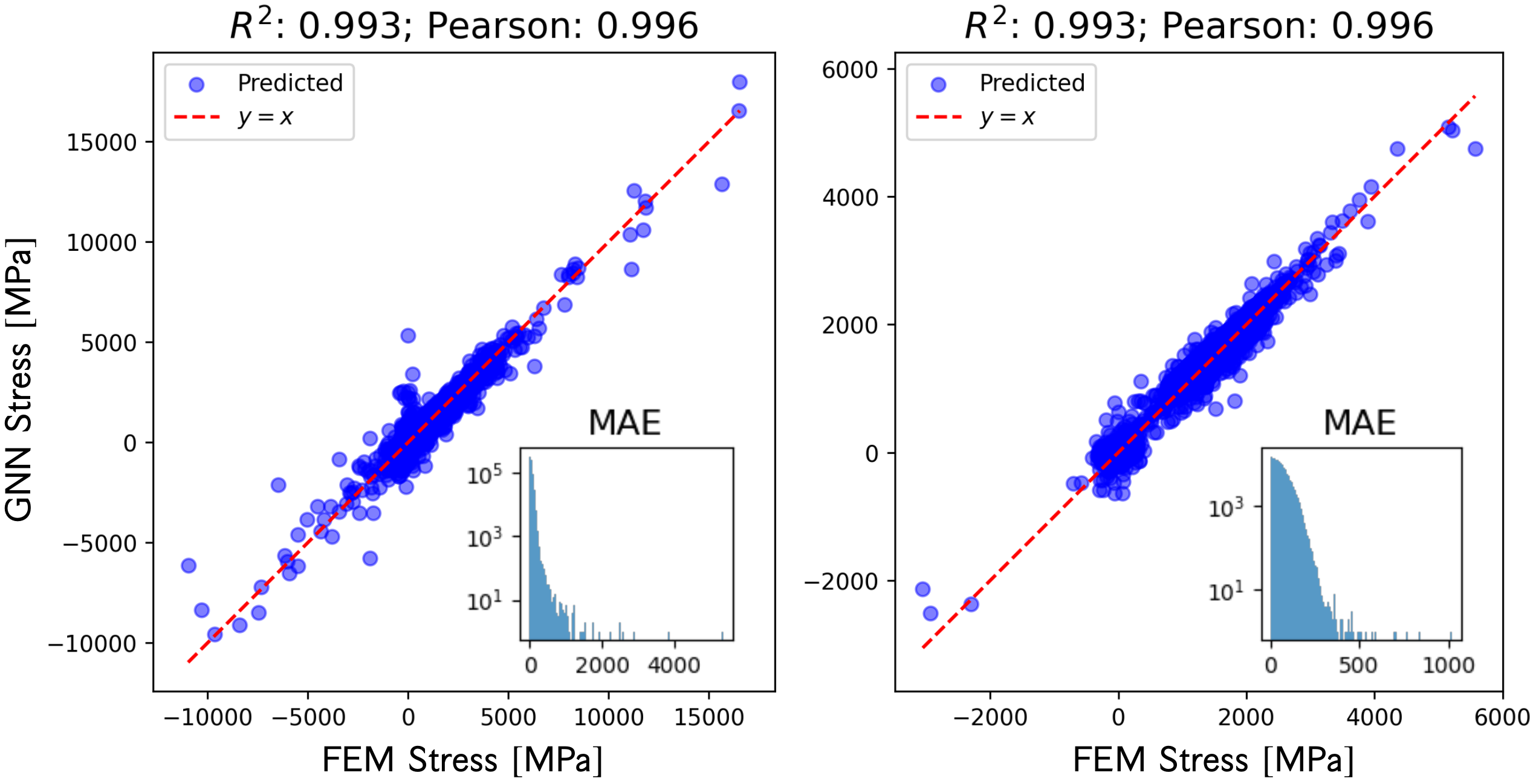}
    \caption{The predictions of the trained model on the training \& testing sets for all the stress components. The left subfigure is the prediction on the training set and the right subfigure is on the testing set. The distribution of MAE is then visualized in the right-bottom corners.}
    \label{fig 3 train & test}
\end{figure}

Figure \ref{fig 3 train & test} displays the overall results of the model training and testing. The left subfigure shows the prediction evaluations on the training set, while the right subfigure shows the prediction evaluations on the testing set. Both subfigures illustrate a high degree of correlation between the predicted and benchmark values, with $R^2$ values of 0.993 and Pearson correlation coefficients of 0.996. The red dashed lines represent the ideal ``$y = x$'' line, indicating predictions equal to benchmark values. The insets in each subfigure show the mean absolute error (MAE) distributions, further highlighting the model's performance. These results demonstrate the model's robust predictive accuracy on both training and testing datasets.

\subsection{Analysis on finite element meshes}

\begin{figure}[htbp]
    \centering
    \includegraphics[width=.85\linewidth]{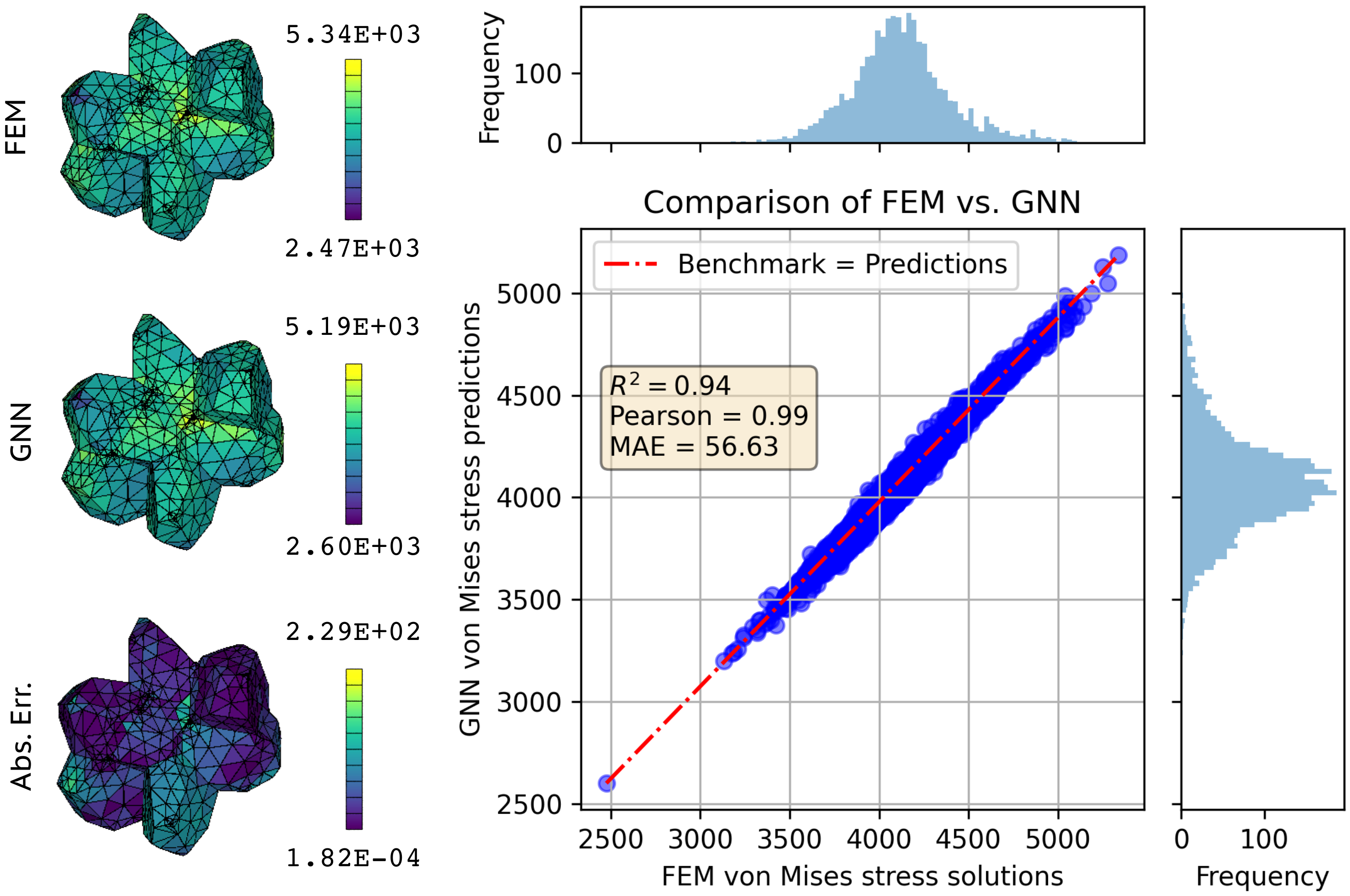}
    \caption{The comparison between FEM and GNN predictions, with absolute errors (visualized on elements with marked color bars) on von Mises stress. The unit for stress is [MPa]. The evaluation of the prediction quality on von Mises stress for the example polycrystal. The unit for stress is [MPa].}
    \label{fig 8 R2 evaluation}
\end{figure}

The von Mises stresses are calculated on each cell as\footnote{using the stress notation introduced in Sec. \ref{sec_data_gen_fem}}\begin{equation}
    \sigma_{\rm vM} = \sqrt{
\frac{1}{2} \left[
(\sigma_{1} - \sigma_{4})^2 +
(\sigma_{4} - \sigma_{6})^2 +
(\sigma_{6} - \sigma_{1})^2 +
6(\sigma_{2}^2 + \sigma_{3}^2 + \sigma_{5}^2)
\right]
}\label{vonMises_eqn}
\end{equation} are visualized on the virtual polycrystals (Figure \ref{fig 8 R2 evaluation}) comparing FEM and GNN, accompanied by the absolute errors. The general stress distribution trends are well learned on the meshes, demonstrated by the stress data distribution. Figure \ref{fig 8 R2 evaluation} quantitatively verifies this observation with a high $R^2$ value of 0.94 and Pearson coefficient of 0.99. The overall MAE for the von Mises stress for this polycrystal is 56.63 [MPa], verifying and quantifying the low deviation of the GNN predictions from the benchmark. Combined analysis from Figure \ref{fig 6 stress components} \& \ref{fig 8 R2 evaluation} detailedly illustrates GNN's effective learning. Note that because the GNN is trained directly on the stress tensor components, accurate predictions of the von Mises stress are inherently expected due to the model's efficient learning capabilities.





\begin{figure}[htbp]
    \centering
    \includegraphics[width=.5\linewidth]{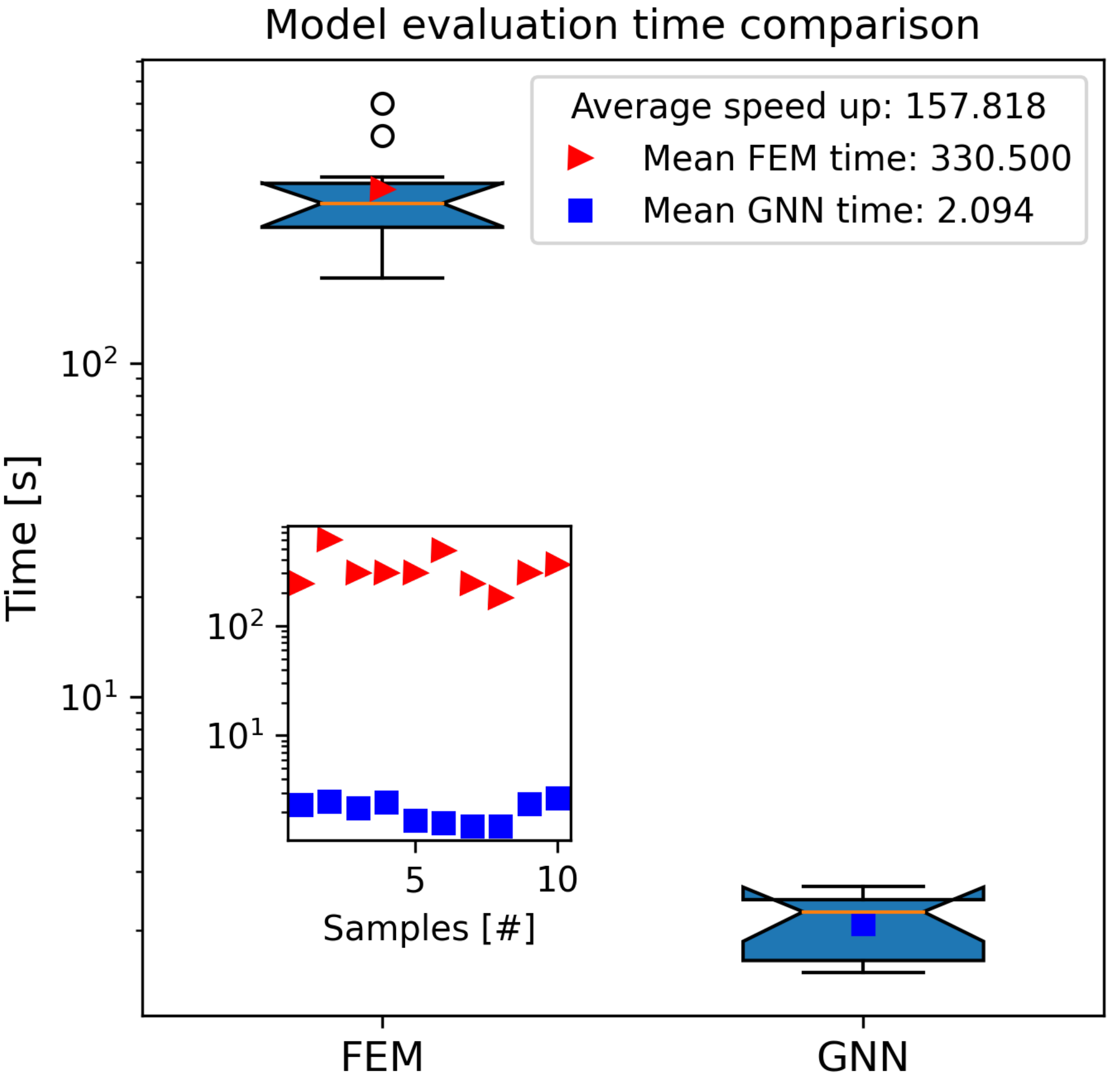}
    \caption{Speed up time comparing constitutive model evaluations of FEM and GNN from 10 randomly selected polycrystal samples. The models are evaluated on a single CPU node on the Sherlock system \cite{stanford_sherlock_docs}.}
    \label{spped up eval}
\end{figure}

One of the main advantages of the proposed approach is that it reduces the computational burden for plasticity modeling. Figure \ref{spped up eval} presents the speed-up evaluation comparing the GNN and FEM methods by comparing the FEM and GNN computational time on 10 randomly selected polycrystals in the testing sets. From the subfigure, one observes that the time does not vary much for the 10 polycrystal samples (blue \& red dots). The average speed-up is estimated at 158, showing that the proposed GNN plasticity can significantly accelerate plasticity modeling with high-accuracy predictions. Several reasons could contribute to this speed-up: (i) In the FEM model, the solver updates stress fields iteratively for each step. This involves nested loops to account for the nonlinear plasticity model, resulting in a significantly increased computational load \cite{neper_cmame_ppr}. (ii) The forward evaluation is computationally efficient in PyTorch \cite{torch_autograd}. (iii) Our model size is compact (Eqn.~\eqref{eqn_fem_stress} with small embedding size); this lightweight nature further contributes to the high-speed evaluation mentioned in (ii).

%

\subsection{Deployment on validation dataset}

\begin{figure}[htbp]
    \centering
    \includegraphics[width=.9\linewidth]{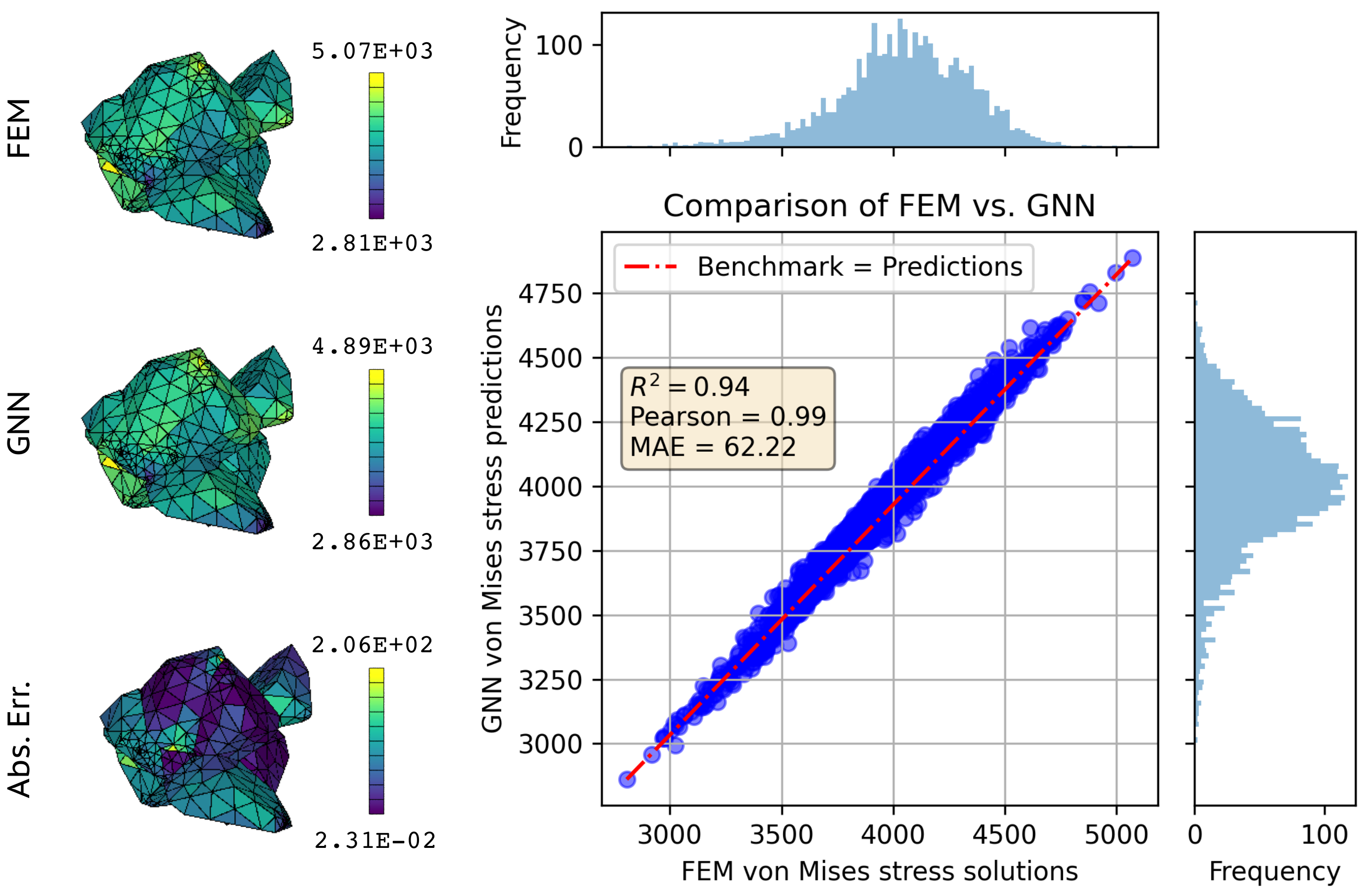}
    \caption{The evaluation of the prediction quality on von Mises stress for an example polycrystal in the validation dataset. The left figures visualize the comparison between FEM and GNN predicted von Mises stress and absolute errors (visualized on elements with color bars marked). The right figure shows the direct map between FEM and GNN predicted von Mises stresses.}
    \label{fig 13 unseen test 1}
\end{figure}

To thoroughly analyze the generalizability of the proposed GNN method, we extend our evaluation beyond the testing sets by running 30 unseen simulations with newly generated polycrystals as the validation set and estimating the prediction quality of the GNN. Figure \ref{fig 13 unseen test 1} provides an analysis of another polycrystal, achieving an overall $R^2$ value of 0.993 for stress components. The von Mises stresses are predicted with high accuracy, as demonstrated by the qualitative observations on the left and quantitative comparisons on the right, yielding $R^2$ values of 0.94 and 0.96. It demonstrates the GNN-based plasticity method generalizes well beyond the training and testing sets, maintaining high-quality predictions on unseen polycrystals. Notably, the polycrystal meshes used in the training, testing, and validation datasets have varying dimensions. Conducting inference on such samples would be nearly impossible with traditional regression methods like vanilla MLPs or CNNs, highlighting the effectiveness of the GNN approach. The overall $R^2$ score for the validation set is 0.992 and a Pearson coefficient of 0.996 (see Figure \ref{fig 13 unseen test 1}). 

\begin{figure}[htbp]
    \centering
    \includegraphics[width=.95\linewidth]{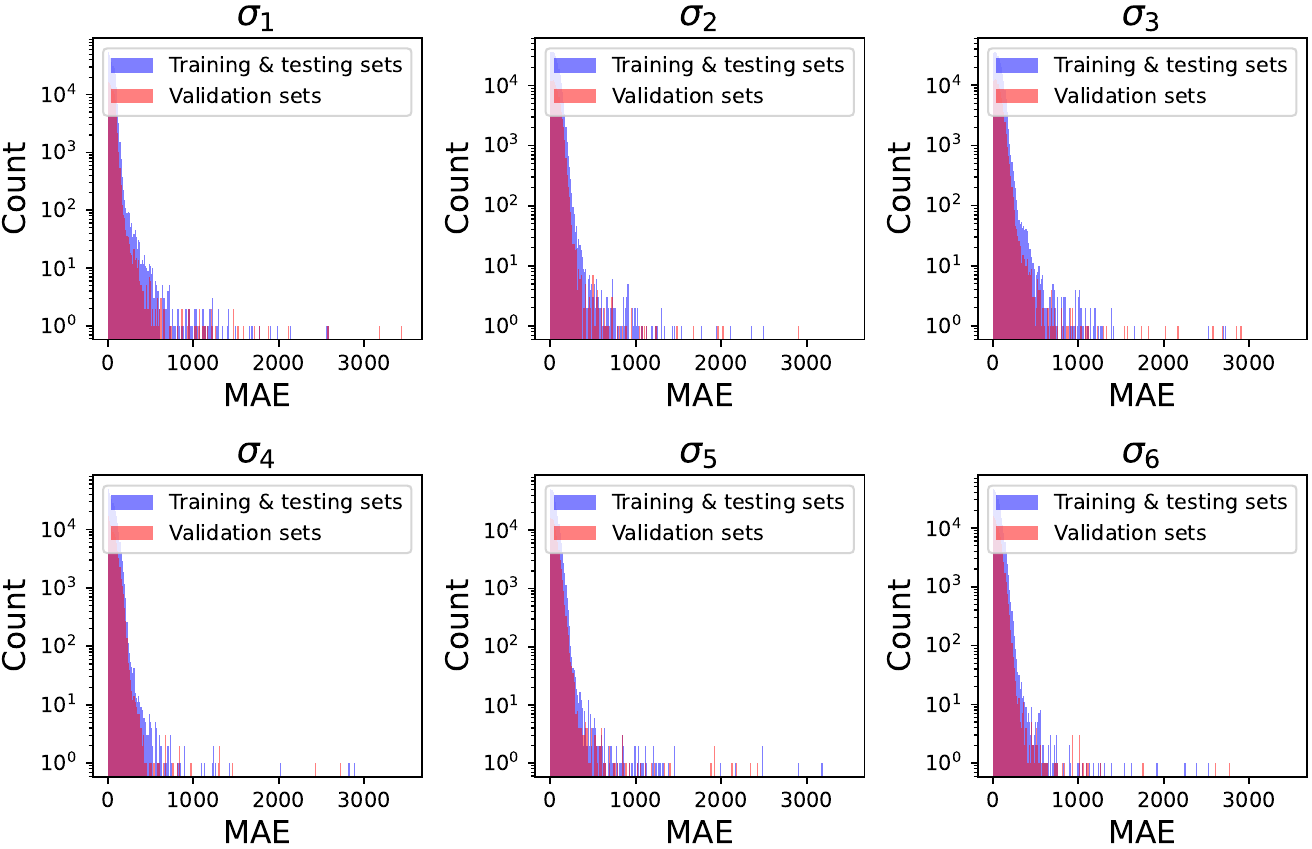}
    \caption{Comparison of absolute error distributions between the training, testing, and validation datasets. }
    \label{fig 15 error analysis}
\end{figure}

Figure \ref{fig 15 error analysis} presents an error analysis that compares training and testing sets with validation data sets in all stress components ($\sigma_1$ to $\sigma_6$), by directly visualizing the MAE distribution. The distributions for the training and testing sets closely resemble those of the validation datasets, validating that the proposed GNN plasticity method does not overfit and generalizes well to the stress distributions across various polycrystals.


\subsection{Limitations of the proposed method}

With the demonstrated fast, accurate, and generalizable predictability of our method on polycrystal plasticity, there are still several limitations. (1) The model is not able to predict the loading path, instead, we are demonstrating the map between strain-stress snapshots that are learnable from our GNN model. (2) Currently, the model is not agnostic to the given material parameters (i.e., elastic moduli, hardening coefficients, etc.)\footnote{according to the data is generated for a defined material in Sec. \ref{sec_data_gen_fem}}. The test cases are material-dependent. (3) Uncoupled stress components with the loading orientation are not accurately captured. By taking all the stress components as a full dataset, since loading coupled stress components $\left\{\sigma_{ij}\ |\ i=1,\ j=1,2,3\right\}$ (using general tensor notation) and uncoupled stress components $\left\{\sigma_{ij}\ |\ i\neq1,\ j=1,2,3\right\}$ are not on the same scale, it is quite challenging for the GNN to accurately predict all the stress components. These are valuable future directions that continue with our current model.


\section{Summary and conclusions\label{sec_conclusion}}

In this paper, we introduce a novel approach for stress predictions using graph neural networks with subgraph training in polycrystal plasticity. The key advantages of our method are: (1) Handling data with varying dimensions --- our GNN model accommodates different node counts generated from various polycrystal meshes, allowing for flexible input data; (2) Efficient subgraph training --- by randomly sampling subgraphs from polycrystals containing $\sim10^5$ nodes and edges\footnote{i.e., on the order of}, we reduce the computing memory requirements; (3) Preserving geometric features --- our GNN model incorporates nodal and edge information, preserving the spatial distribution of stress and strain, thereby enhancing the ``learnability" of the data.

Our numerical experiments demonstrate that the GNN model accurately predicts stress components, achieving $R^2$ scores greater than 0.99 on the training, testing, and validation datasets. Additionally, the von Mises stress predictions for the polycrystals indicate that the proposed GNN method accurately captures von Mises stress features. The model generalizes well beyond the training and testing data, as evidenced by the similar MAE distribution across the training, testing, and validation datasets. The proposed GNN method speeds up stress predictions in the plastic regime more than 150 times compared with the benchmark finite element methods.

We also briefly outline the limitations of our framework: stress components that are uncoupled from the loading direction are not accurately captured. Only the map between stress-strain snapshots is the learning target for our GNN. However, these uncoupled stress components will not evidently affect the effectiveness of the GNN method in mechanical analysis, particularly when estimating the critical stress under plastic deformation as the von Mises stresses are accurately captured.

This work outlooks surrogate modeling of polycrystal plasticity using graphs to demonstrate the transient strain-stress map can be learned by GNNs. Future work could include incorporating physics-informed features into the framework and tackling more path-dependent plasticity modeling tasks, leaving open space to the field.



\section*{Data Availability}


The associated codes and data are available at \url{https://gitlab.com/hanfengzhai2/GNN-FEM-PolyPlas}. All data and code are published
under \href{https://gitlab.com/hanfengzhai2/GNN-FEM-PolyPlas/-/blob/main/LICENSE}{MIT License}. The virtual polycrystals are generated and visualized using {\em Neper}, accessible at \url{https://neper.info/}, The FEM plasticity simulations utilize the open-source software package {\em FEPX}, publicly available at \url{https://fepx.info/index.html} \cite{fepx_arxiv, fepx_original}.

\section*{Conflict of interest}

None.

\section*{Acknowledgment}

The author acknowledges support from the Enlight Foundation Graduate Fellowship via Leland Stanford Junior University. The author thanks Myung Chul Kim of Stanford University for discussions on presampling algorithms for training GNNs, Romain Quey of CNRS for discussions on the implementation of {\em Neper} and {\em FEPX} for simulation visualizations, and Matthew Kasemer of The University of Alabama for the general comments on GNN, discussions on plasticity theory, and on the {\em FEPX} implementation. The author also thanks the anonymous reviewers for their invaluable comments, which significantly improved the manuscript.

\appendix
\renewcommand{\thefigure}{A\arabic{figure}} 
\setcounter{figure}{0}  
\renewcommand{\theequation}{A.\arabic{equation}} 
\setcounter{equation}{0}  

\clearpage

\section{Data preparation}

\subsection{Graph conversion methods\label{appendix graph conversion}}

\begin{figure}[htbp]
    \centering
    \includegraphics[width=0.8\linewidth]{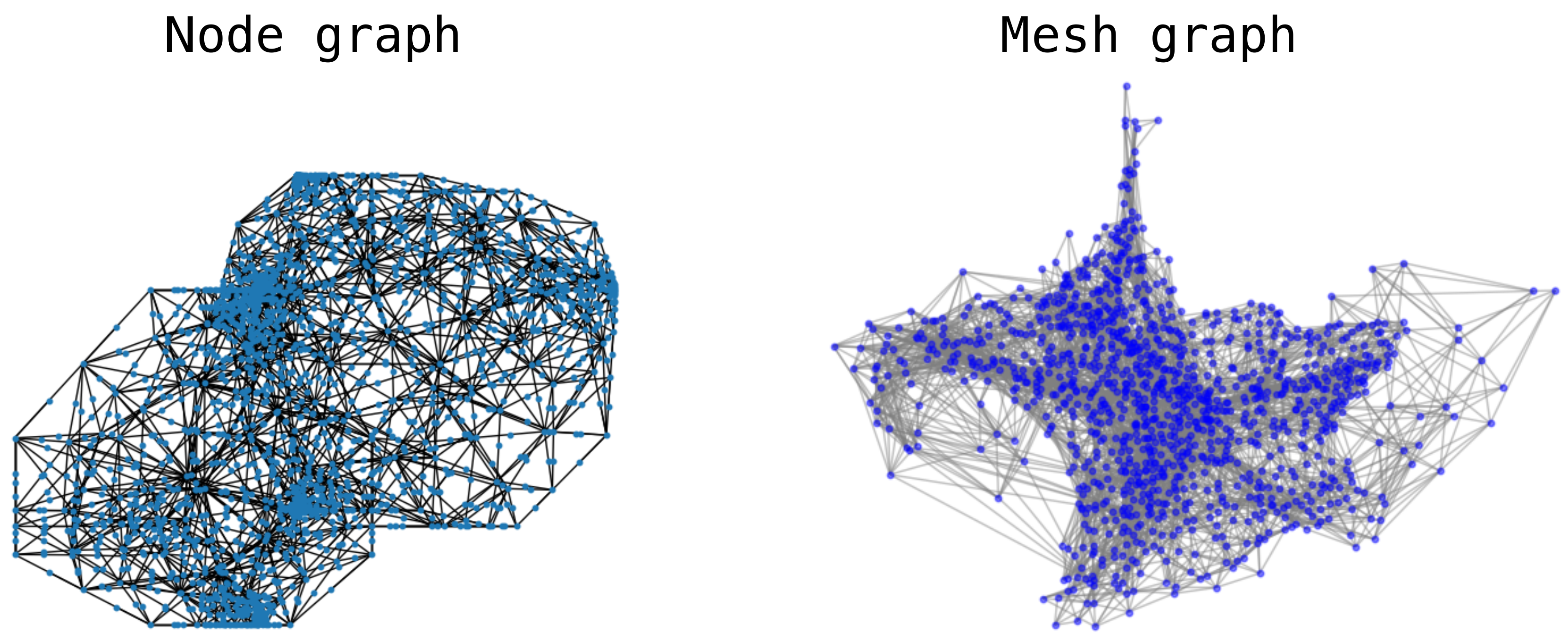}
    \caption{Schematic illustration for different graphs obtained from FEM mesh nodes and cells.}
    \label{node to mesh graph}
\end{figure}

The training graphs were converted from the FEM mesh cells and the neighboring connections (Sect.~\ref{fig_mesh_graph}). There were three main considerations for converting the graph in such a way: (1) The numerical values from FEM are solved on mesh cells. Hence, directly converting mesh elements to nodes makes defining the map from strain to stress much more straightforward. (2) Such FEM graphs are agnostic to the order of the test functions used in our FEM calculations. (3) In some sense, only the ``first-order'' connections between the mesh cells are created as edges, making the conversion process much faster and more efficient. The FEM graphs converted from this method are inspired by respecting the conformality of FEM, in which two common mesh elements share the edge. One of the other intuitive ways is to use the FEM node as a graph vertex directly, where nodal connections are edges. The left subfigure of Figure \ref{node to mesh graph} (Node graph) illustrates this graph conversion method. As mentioned previously, the drawbacks are that it is hard to define the strain-stress map on the graphs, that elements of higher order introduce redundant nodes, and that the edges do not preserve the connections between element cells. The right subfigure of Figure \ref{node to mesh graph} (Mesh graph) illustrates the graph using the mesh cell method we proposed.


\subsection{Finite element implementation of crystal plasticity\label{appendix fem alg}}

Here we briefly discuss the implementation of crystal plasticity using FEM using {\em FEPX}. Note that only the main steps are summarized herein for a clearer understanding of the manuscript; for numerical implementation details, please refer to Refs.~\cite{fepx_original, fepx_arxiv}. For the elastic deformation, the cubic symmetry for the stiffness matrix is employed, representing the stress-strain relationship following:\begin{equation}
    \begin{aligned}
        \begin{bmatrix}
            \tau_{11}\\\tau_{22}\\\tau_{33}\\\tau_{23}\\\tau_{13}\\\tau_{12}
        \end{bmatrix}=\begin{bmatrix}
            \mathcal{C}_{11} & \mathcal{C}_{12} & \mathcal{C}_{12} & & &\\
            \mathcal{C}_{12} & \mathcal{C}_{11} & \mathcal{C}_{12} & & &\\
            \mathcal{C}_{12} & \mathcal{C}_{12} & \mathcal{C}_{11} & & &\\
            &&&\mathcal{C}_{44}&&\\
            &&&&\mathcal{C}_{44}&\\
            &&&&&\mathcal{C}_{44}\\
        \end{bmatrix}\begin{bmatrix}
            e_{11}\\e_{22}\\e_{33}\\2e_{23}\\2e_{13}\\2e_{12}
        \end{bmatrix}
    \end{aligned}
\end{equation}where $e_{ij}$ and $\tau_{ij}$ are the shear strain and stress.

In the kinematic evolution, the motion can be split into volumetric and deviatoric parts\footnote{for convenience of numerical implementation}, in which the elasticity equation relating the Kirchhoff stress and elastic strain writes:\begin{equation}
    \begin{aligned}
        {\tt tr}\{{\bf \tau}\} = \frac{\kappa}{3} {\tt tr} \{\mathbf{e}^e\},\quad \{{\bf \tau'}\} = \left[\mathbb{C}'\right]\{\mathbf{e}^{e'}\}
    \end{aligned}
\end{equation}where $\kappa$ is bulk modulus, following the relationship $\kappa =3(\mathcal{C}_{11}+2\mathcal{C}_{12})$. For the components in $\left[\mathbb{C}'\right]$, they relates to $\left[\mathbb{C}\right]$ as $\mathcal{C}_{11}' =\mathcal{C}_{11}-\mathcal{C}_{12}$,  $\mathcal{C}_{22}' =\mathcal{C}_{11}-\mathcal{C}_{12}$, and $\mathcal{C}_{33}'=\mathcal{C}_{44}'=\mathcal{C}_{55}' = \mathcal{C}_{44}$ based on cubic symmetry.

The spatial time-rate difference of the elastic strain can be expressed in finite difference scheme:\begin{equation}
    \begin{aligned}
        \{\dot{\mathbf{e}}^e\} = \frac{1}{\Delta t} \left(\{\mathbf{e}^e\} - \{\mathbf{e}_0^e\}\right)
    \end{aligned}
\end{equation}where $\{\mathbf{e}^e\}$ and $\{\mathbf{e}_0^e\}$ are the elastic strains at the end and beginning of a time step. $\Delta t$ is the time step.

For the deviatoric portion of the motion, the deformation rate can be expanded in the form (from Equation \eqref{eqn sym and spin tensors})\begin{equation}
    \begin{aligned}
        \{\mathbf{D}'\} = \frac{1}{\Delta t} \left\{\mathbf{e}^{e'}\right\} + \left\{\hat{\bf D}^p\right\} + \left[\hat{\bf W}^p\right] \left\{\mathbf{e}^{e'}\right\} - \frac{1}{\Delta t}\left\{\mathbf{e}_0^{e'}\right\}
    \end{aligned}
\end{equation}

The deviatoric portion of the Cauchy stress can be updated following:\begin{equation}
    \begin{aligned}
        \left\{\sigma'\right\} = \left[\mathbf{s}\right] \left(\left\{\mathbf{D}'\right\} - \{\mathbf{h}\}\right)
    \end{aligned}
\end{equation}where\begin{equation}
    \begin{aligned}
        \left[\mathbf{s}\right]^{-1} = \frac{\beta}{\Delta t} \left[\mathbb{C}'\right]^{-1} + \beta \left[\mathbf{m}\right]\\
        \left\{\mathbf{h}\right\} = \left[\hat{\mathbf{W}}^p\right]\left\{\mathbf{e}^{e'}\right\} - \frac{1}{\Delta t} \left\{\mathbf{e}_0^{e'}\right\}
    \end{aligned}
\end{equation}in which $\beta = {\tt det}(\mathbf{v}^e)$, and $[\mathbf{m}]$ is the map from the deviatoric Kirchhoff stress $\tau'$ to $\hat{\mathbf{D}}^p$.

The interpolation function $\left[ \mathsf{N} (\xi, \eta, \zeta)\right]$ interpolates the nodal coordination points and the velocity fields via $\left\{{\sf x}\right\} = \left[ \mathsf{N} (\xi, \eta, \zeta)\right]\left\{{\sf X}\right\}$ and $\left\{{\sf v}\right\} = \left[ \mathsf{N} (\xi, \eta, \zeta)\right]\left\{{\sf V}\right\}$ for the global assembly. Under the Galerkin formulation, weight functions were used to construct the residual, where the weight function $\psi$ can be interpolated in the same way:\begin{equation}
    \left\{\psi\right\} = \left[ \mathsf{N} (\xi, \eta, \zeta)\right]\left\{\Psi\right\}
\end{equation}

To achieve equilibrium for the system, one would solve the global weighted residual equation:\begin{equation}
    R_u = \int_\mathcal{B} \psi \cdot \left(\nabla\cdot \mathbf{\sigma} + 
 \mathbf{f}\right)d\mathcal{B} = 0
\end{equation}where $\mathcal{B}$ is the continuum body to be solved.

After the global assembly, one could solve the nonlinear system to obtain the velocity field $\mathsf{V}$ from\footnote{nonlinear in the sense that $\left[\mathsf{K}_d\right]$ and $\left[\mathsf{K}_v\right]$ depends on $[\mathsf{V}]$}:\begin{equation}
    \left[\left[\mathsf{K}_d\right] + \left[\mathsf{K}_v\right]\right]\left\{\mathsf{V}\right\} = \left\{\mathsf{F}_a\right\} + \left\{\mathsf{F}_d\right\} + \left\{\mathsf{F}_v\right\}
\end{equation}where the details of calculating the elements in the global stiffness matrix and force vectors can be checked in Ref.~\cite{fepx_arxiv}.

\section{Graph neural networks\label{appendix_gnn}}

\subsection{Training procedure\label{appendix training}}

\begin{figure}[htbp] \centering \includegraphics[width=0.65\linewidth]{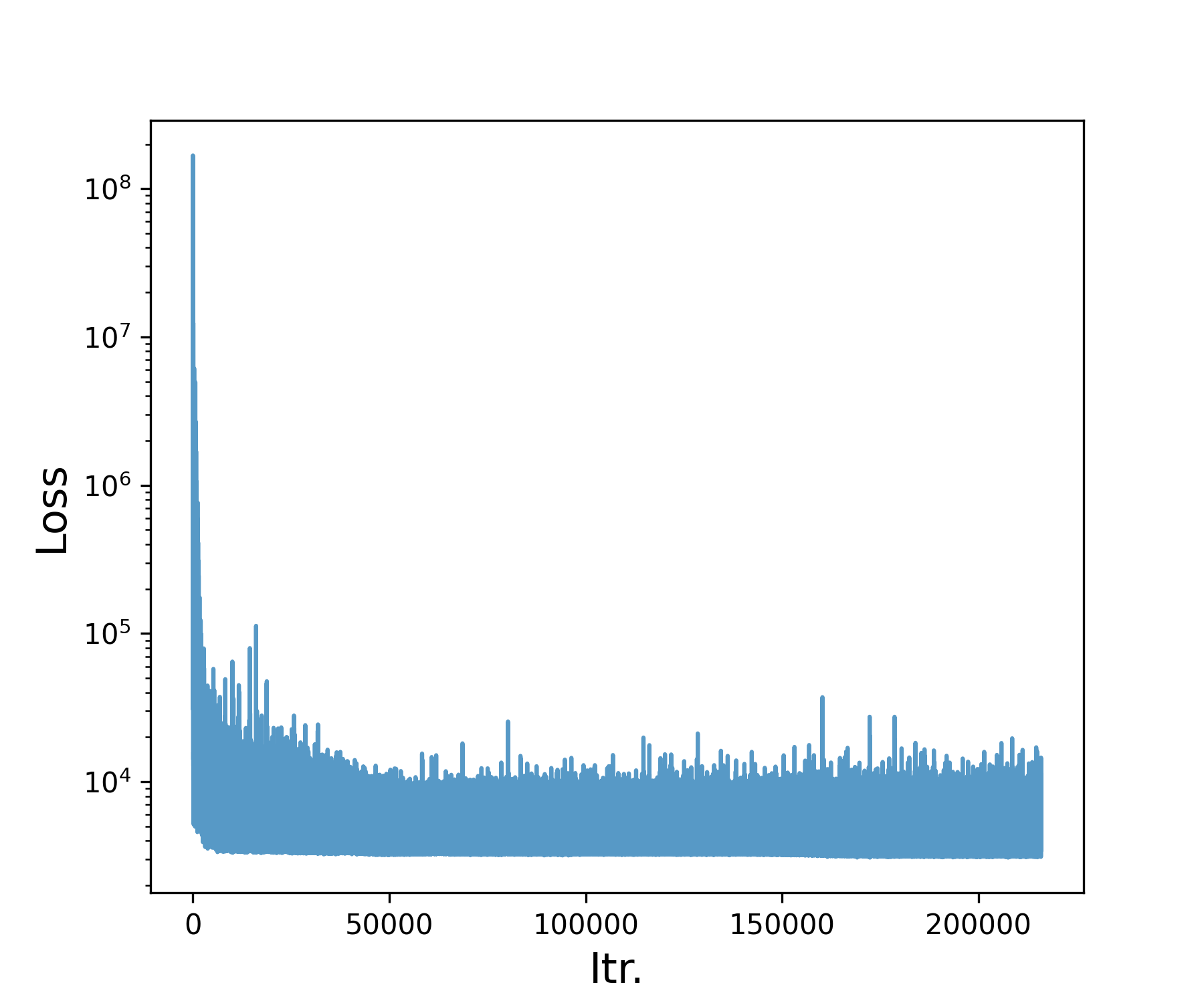} \caption{Loss curve w.r.t. iterations during the training process for 3000 epochs.} \label{fig:epoch charac} \end{figure}

Since the model is directly trained on the stress data ($\sim 10^3$), the loss is the MSE loss between the predicted \& benchmark stress, hence displaying a large loss also on the scale of $10^3$. Based on our observations, the training loss shows a minimal decrease beyond 1000 epochs, indicating that the model has effectively converged. To further investigate this, we extended the training to 3000 epochs, and the corresponding loss curve is presented in Figure \ref{fig:epoch charac}.

It is important to note that the GNN is iteratively trained on 72 graphs for each epoch. Thus, training for 3000 epochs corresponds to approximately 216,000 iterations. The figure shows that the loss stabilizes after approximately 5000 iterations, plateauing to fluctuating values without significant further reduction. This behavior suggests that 1000 epochs are sufficient for the GNN to achieve ``convergence''.

We believe this demonstrates that the choice of 1000 epochs is appropriate for training the GNN, as it ensures convergence without unnecessary computational overhead.








\subsection{Model characterization}

\begin{figure}[htbp]
    \centering
    \includegraphics[width=0.5\linewidth]{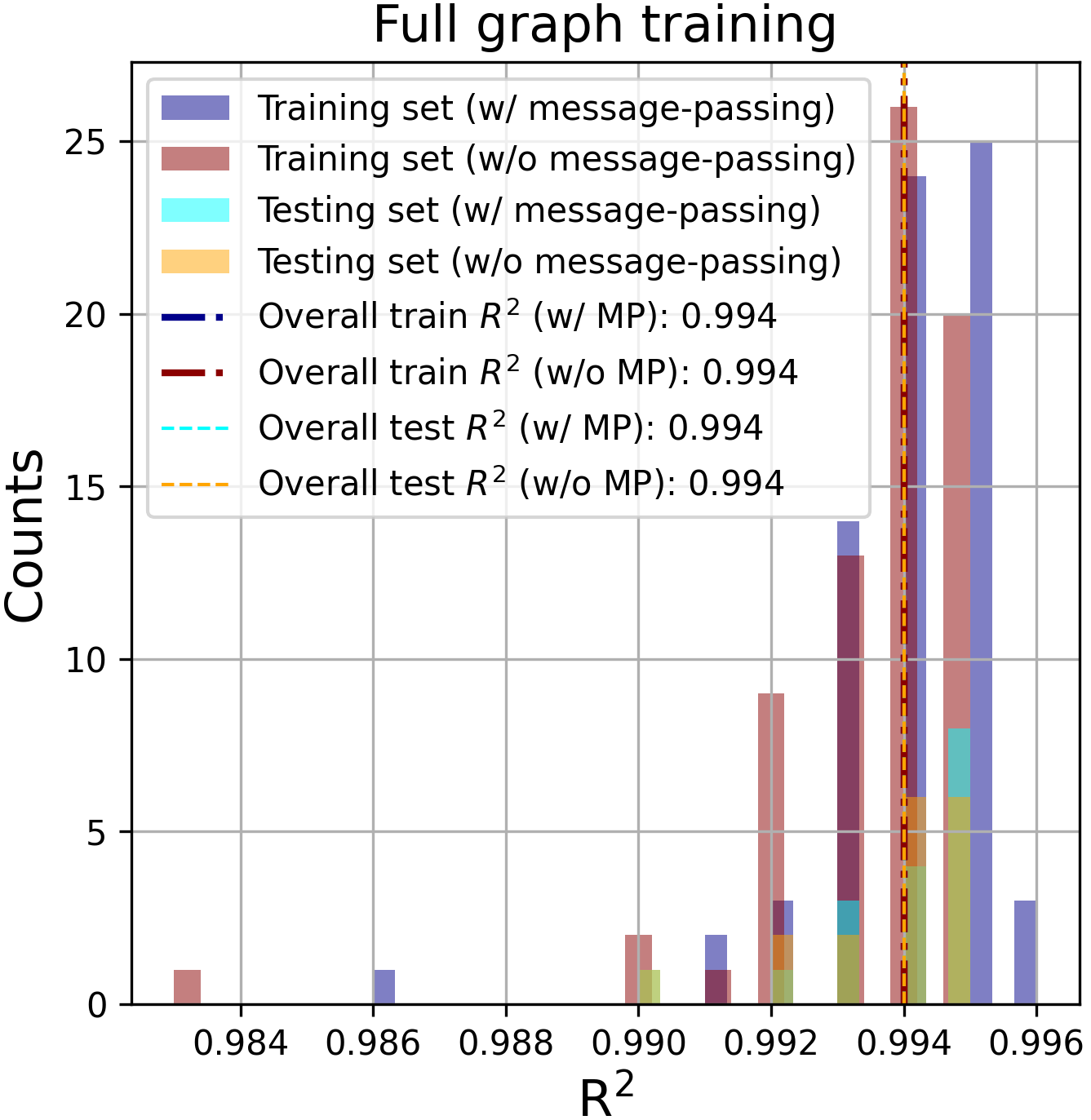}
    \caption{$R^2$ values comparing GNN with and without message-passing mechanisms for full-graph training.}
    \label{fig:mess pass charac}
\end{figure}

Figure~\ref{fig:mess pass charac} shows the model prediction quality with and without message-passing (MP) layers tested with full graph ratio. Interestingly, for this test, both models perform well with high-quality predictions. However, it can be seen that the predictions with MP have more high-quality predictions by comparing the blue and red bars and comparing the light blue with the yellow bars at the relatively lower $R^2$ scores regime ($<0.99$). This suggests that the message-passing mechanism improves GNN's predictability but does not significantly improve the model.

\begin{figure}[htbp]
    \centering
    (a)\includegraphics[width=0.43\linewidth]{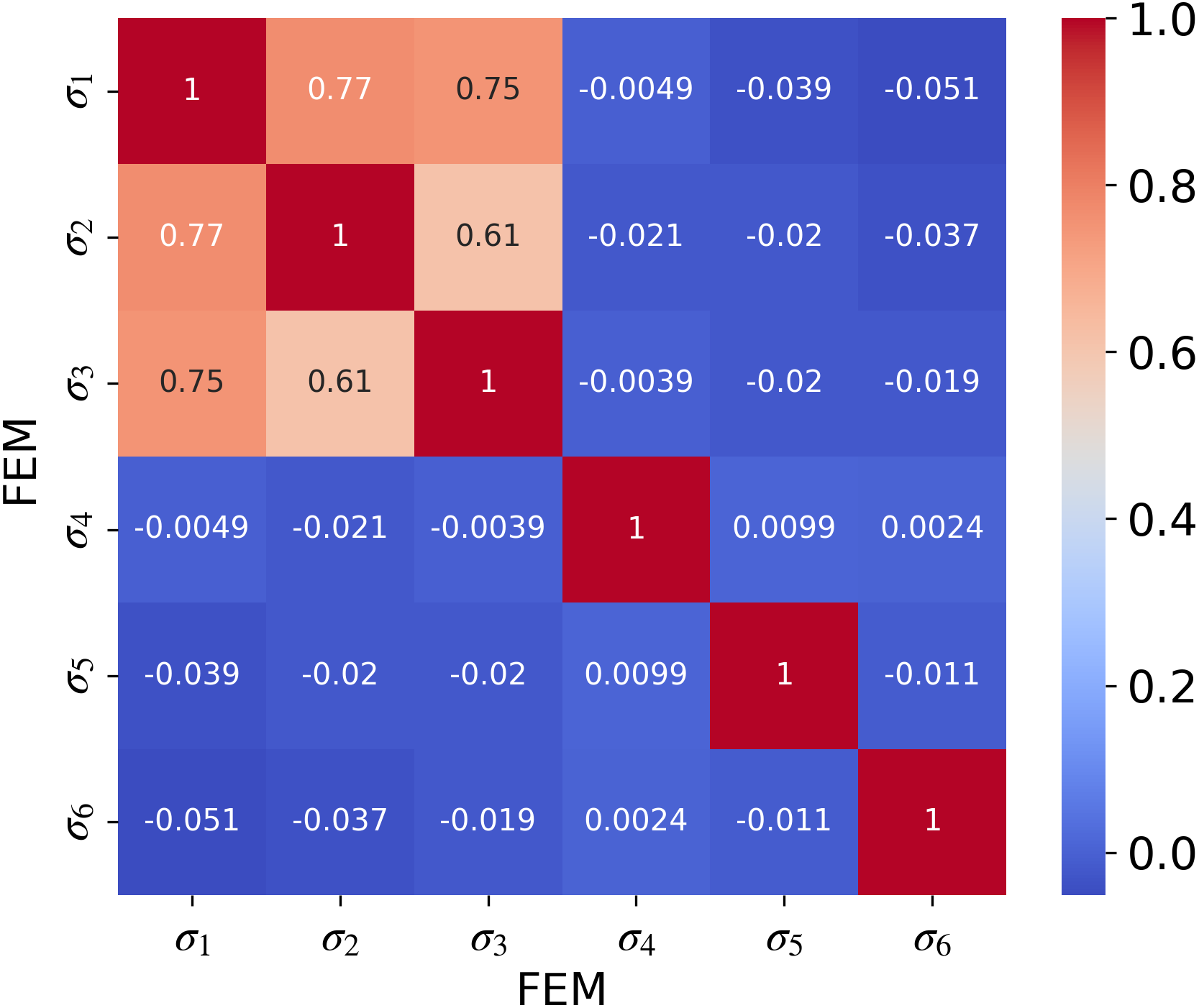}\hspace{15pt}(b)\includegraphics[width=0.43\linewidth]{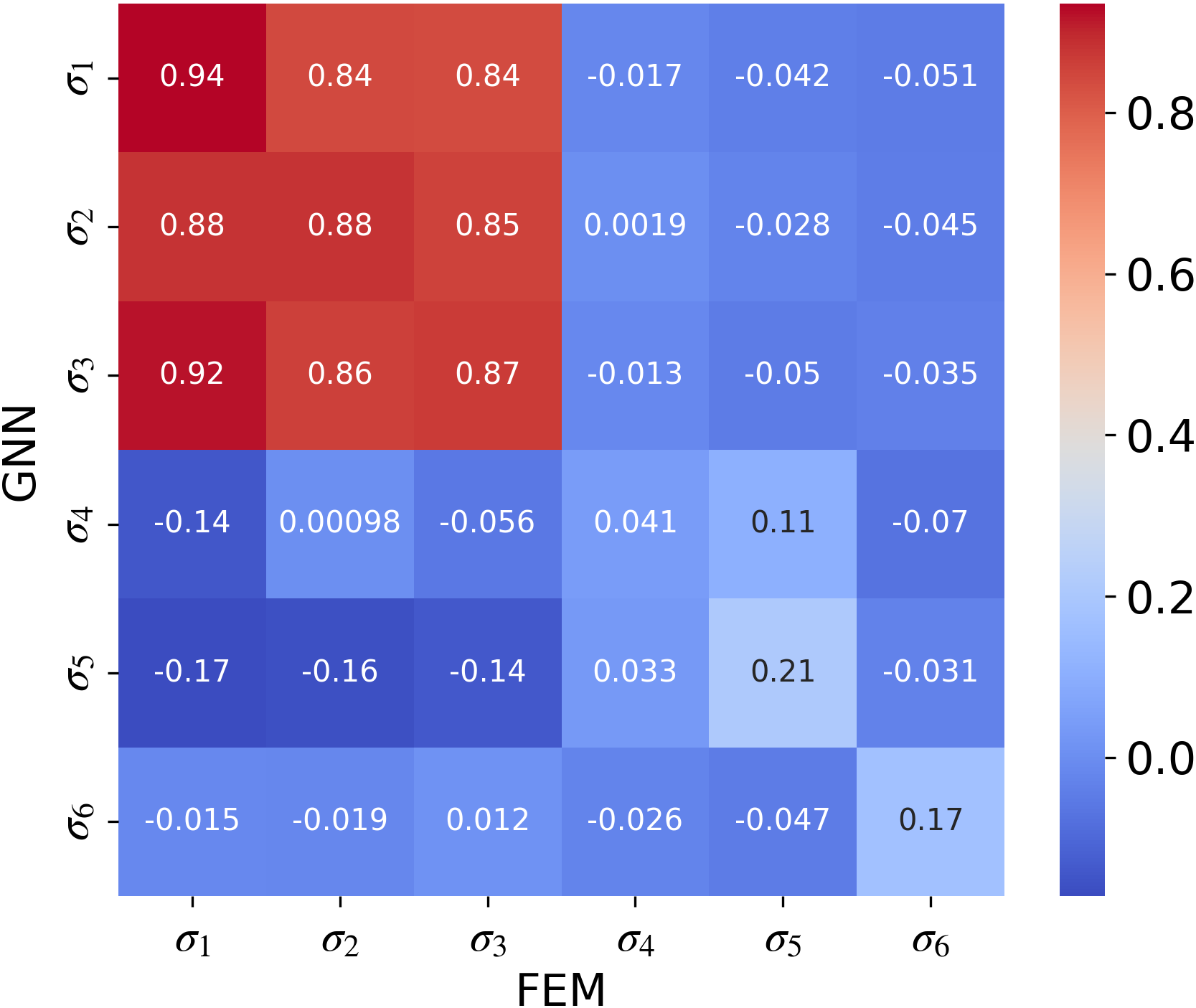}
    \caption{Pearson correlation between the stress components. (a) Self-correlations for the FEM calculated stress components. (b) Correlations for calculated stress components predicted by GNN and FEM.}
    \label{fig:stress components pearson}
\end{figure}

Figure \ref{fig:stress components pearson} (a) visualizes the Pearson correlations between the FEM calculated stress components, indicating the correlations between $\sigma_{1j},\ j=1,2,3$ is much higher than that of other stress components. Figure \ref{fig:stress components pearson} (b) visualizes the Pearson correlations between the stress components predicted by GNN and FEM. It can be observed that GNN predicts much more accurately for stress components $\sigma_{1j}$. This suggests that the loading-coupled stress components, whose both numerical values and correlations are higher, are much more well-predicted by the GNN, which is consistent with the physical scenario in which 1-directional loading is applied.

\begin{figure}[htbp]
    \centering
    \includegraphics[width=0.5\linewidth]{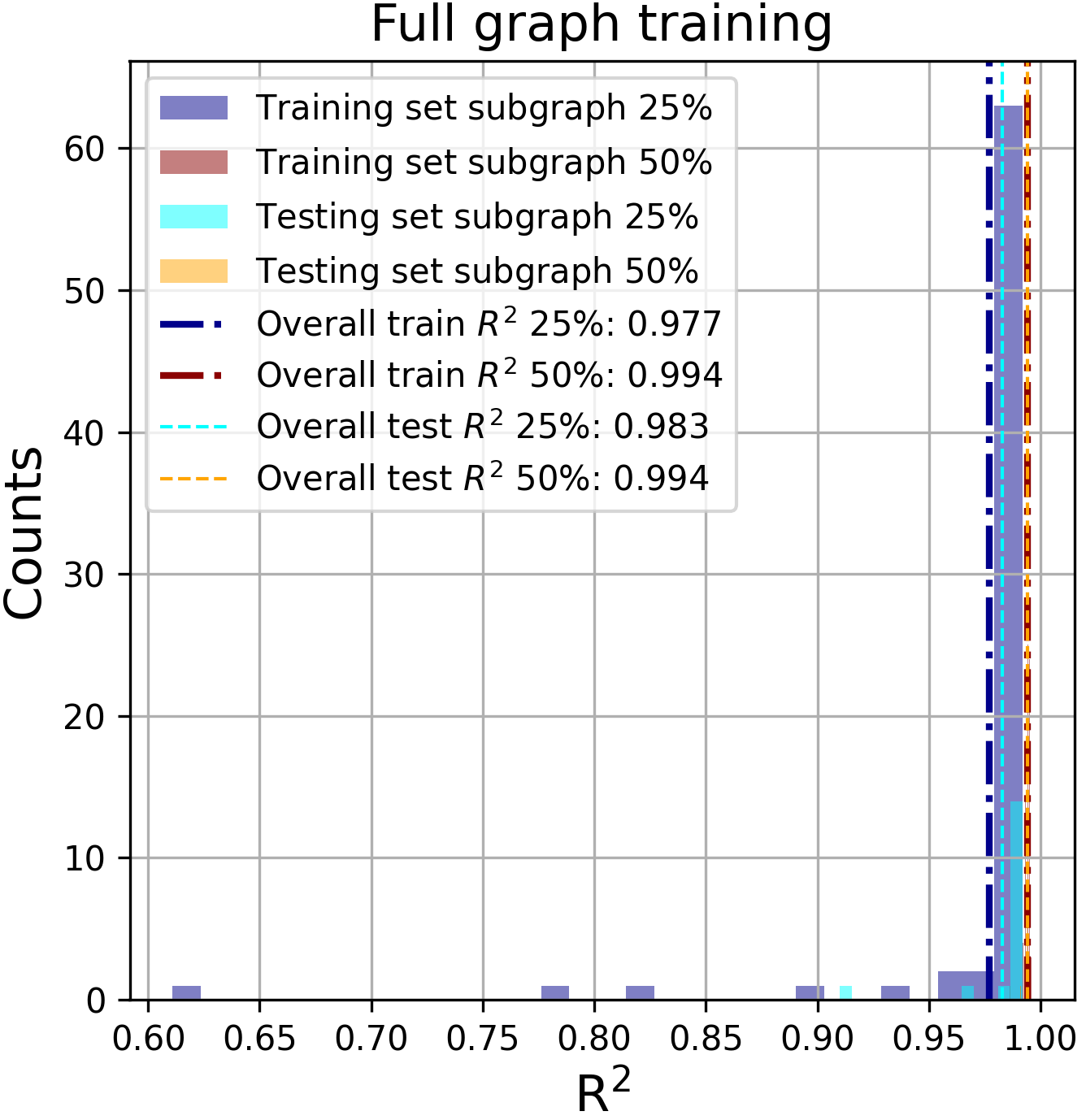}
    \caption{$R^2$ scores from GNN predictions based on different subgraph ratios.}
    \label{fig:subgraph ratio}
\end{figure}

Figure \ref{fig:subgraph ratio} presents the $R^2$ scores for GNN predictions trained with varying subgraph ratios. The results indicate that a subgraph ratio of $0.25$ does not yield accurate predictions. However, increasing the ratio to $0.5$ allows the GNN to achieve an overall $R^2 = 0.994$ for both training and testing sets, with no low-accuracy samples observed. Per our standard, $R^2 > 0.99$ signifies a high-accuracy model. These findings suggest that a subgraph ratio of $0.5$ is a decent balance, enabling the GNN to maintain high accuracy while preserving the essential information of the graph.

\section{Additional results}

\subsection{Predictions from GNN}

\begin{figure}[htbp]
    \centering
    \includegraphics[width=.9\linewidth]{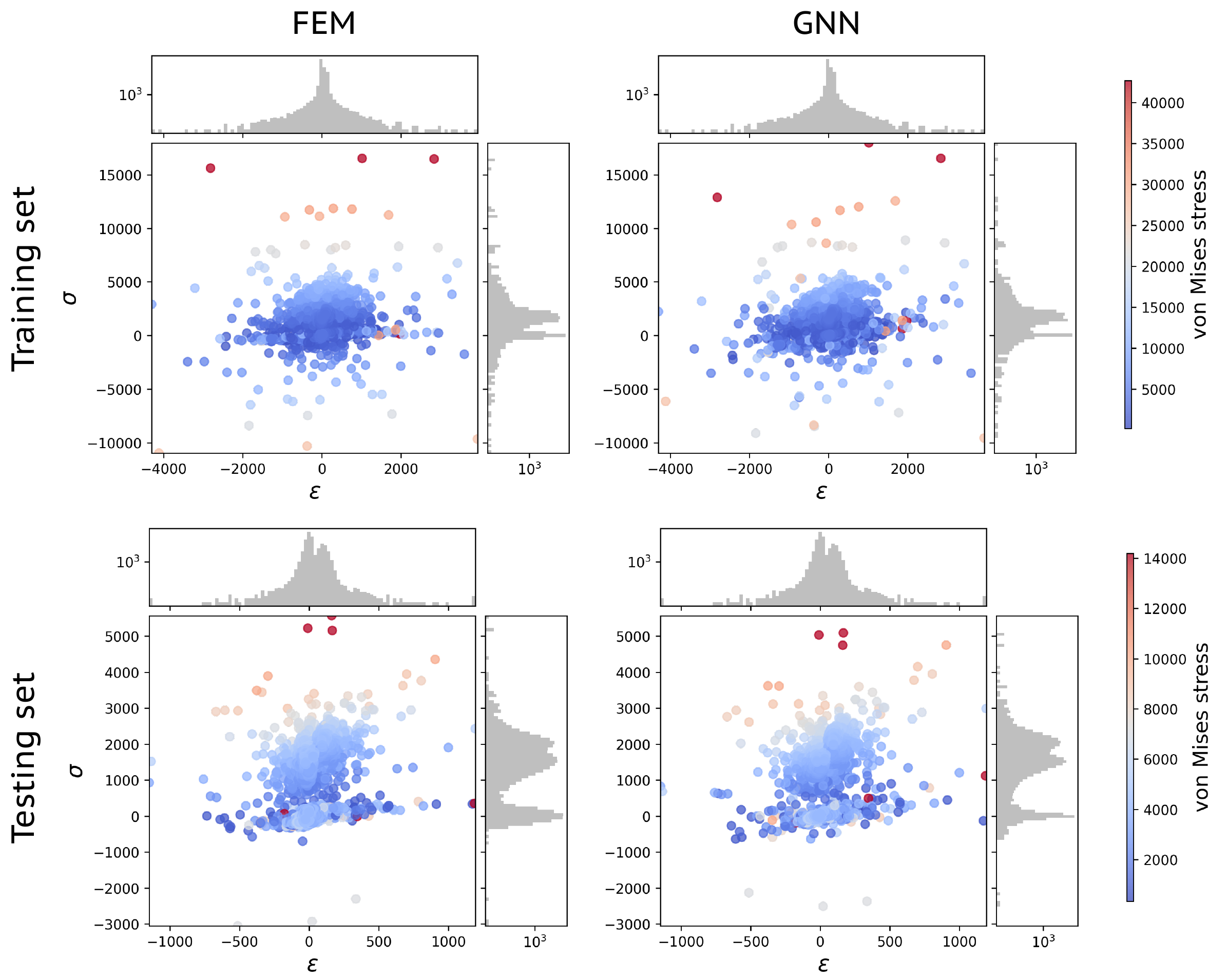}
    \caption{The comparison between the ground truth (by {\em FEPX}) and predictions on the strain-stress maps for all the stress \& strain components. The upper figures correspond to the training sets.  The bottom figures correspond to the testing sets. The data points are visualized according to the calculated von Mises stresses.}
    \label{fig 5 epsilon & sigma11}
\end{figure}

Accompanying the high $R^2$ values, to directly verify the high-quality predictions using GNN, the stress values on each finite element cell for both the training and testing sets are visualized (Figure \ref{fig 5 epsilon & sigma11}). The general data trends are well captured. With von Mises stress as label marks, the predictions preserve the stress distribution among mesh cells for both the training and testing sets, indicating no overfitting for the proposed method. Figure \ref{fig 3 train & test} demonstrates the quality of the predictions with direction prediction data visualization and high $R^2$ scores.


\begin{figure}[htbp]
    \centering
    \includegraphics[width=.8\linewidth]{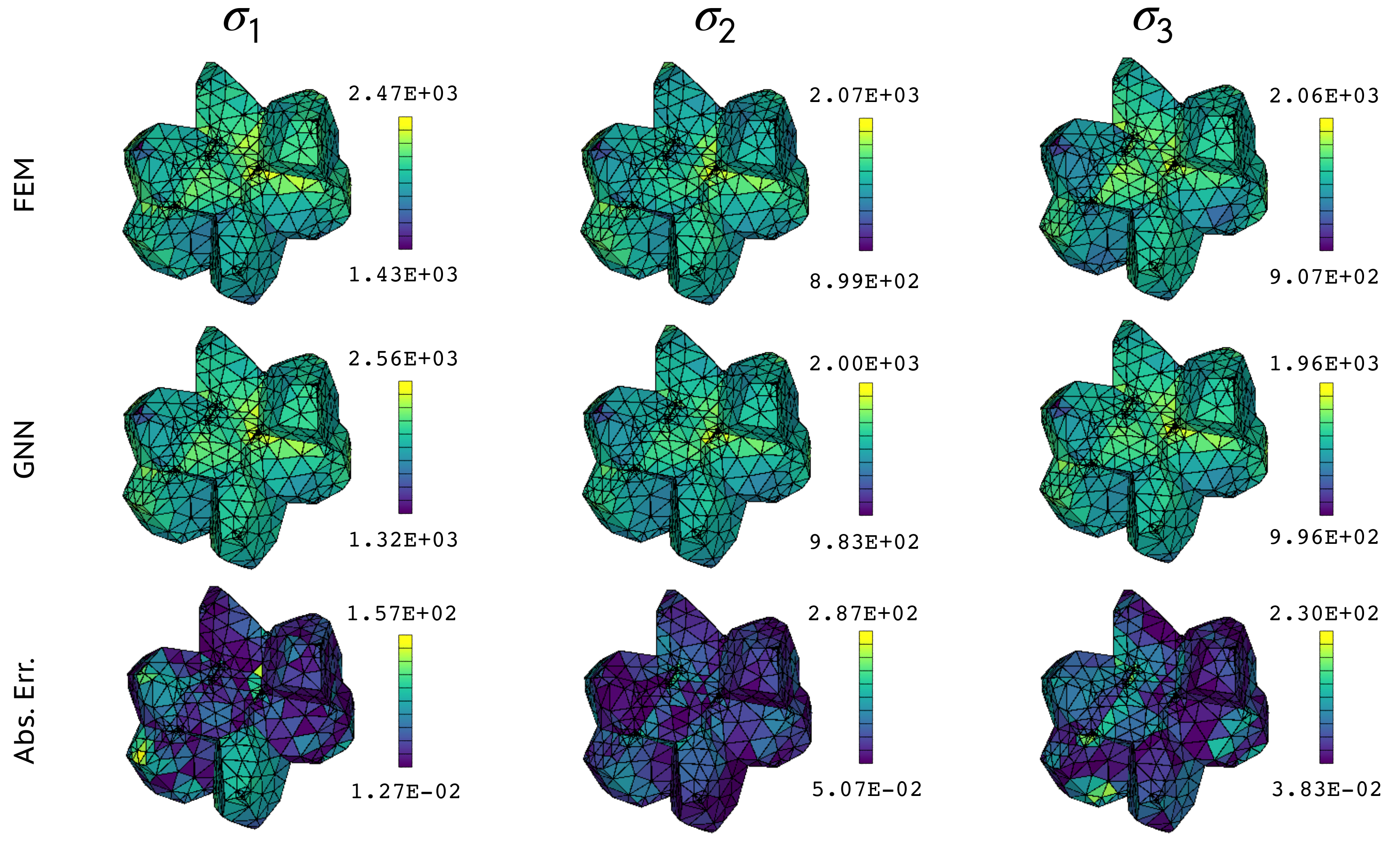}
    \caption{The comparison between FEM and GNN predictions, with absolute errors of stress components on one example grain in the testing sets for stress components $\sigma_1$, $\sigma_2$ and $\sigma_3$ (visualized on elements with marked color bars). The unit for stress is [MPa].}
    \label{fig 6 stress components}
\end{figure}

Associated with the visualization of the data (Figure \ref{fig 5 epsilon & sigma11}), we directly visualize the stress distributions on the virtual polycrystals (Figures \ref{fig 6 stress components}) for a test polycrystal with $R^2$ of 0.994. Data scales are well captured by the GNN for each stress component $\sigma_i$, with comparably small absolute errors (Figure \ref{fig 6 stress components}). GNN predicts much more accurately in the active loading directions by learning different stress value ranges for each component. 

\begin{figure}[htbp]
    \centering
    \includegraphics[width=\linewidth]{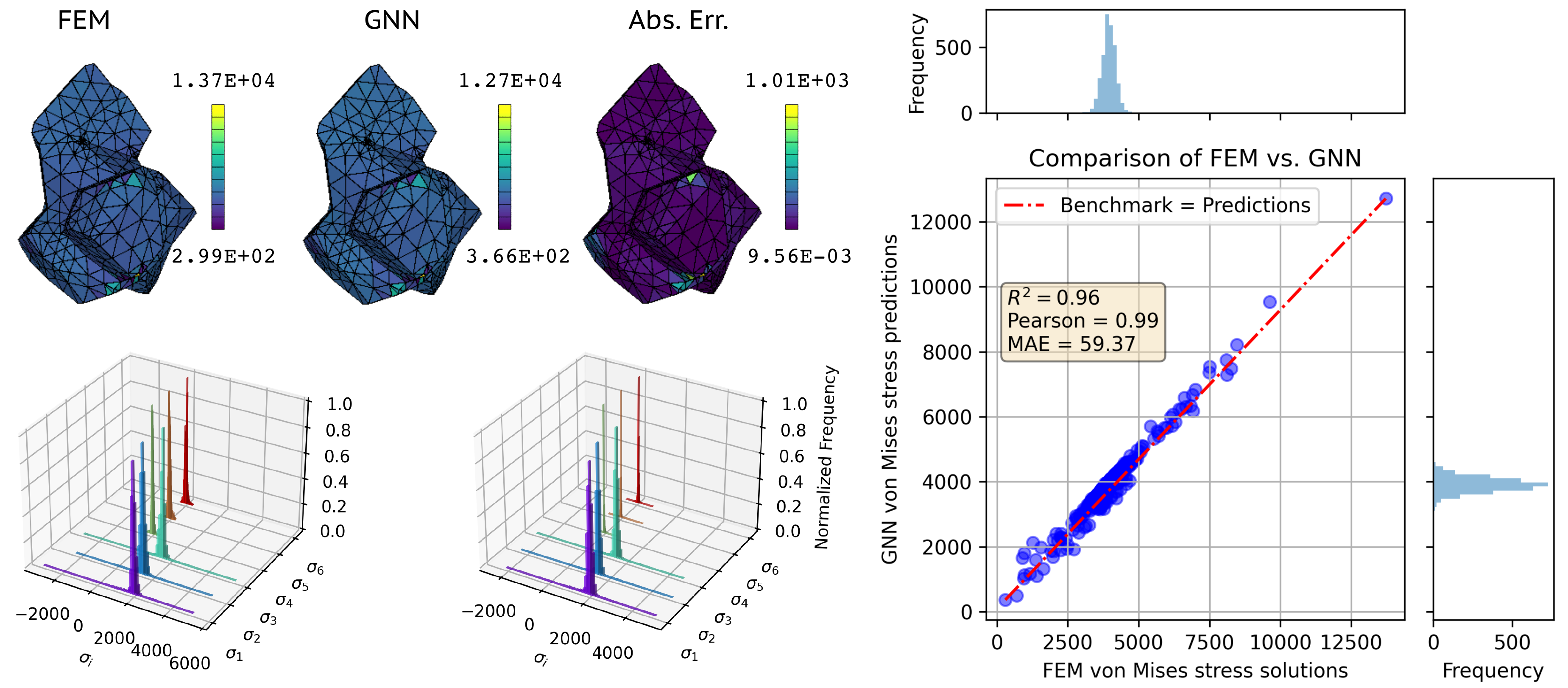}
    \caption{The evaluation of the prediction quality on von Mises stress for another example polycrystal. The upper left figures visualize the comparison between FEM and GNN-predicted von Mises stress and absolute errors (visualized on elements with color bars marked). The bottom left figures visualize the distributions of different stress components. The right figure shows the direct map between FEM and GNN predicted von Mises stresses.}
    \label{fig 9 test 2}
\end{figure}

\begin{figure}[htbp]
    \centering
    \includegraphics[width=\linewidth]{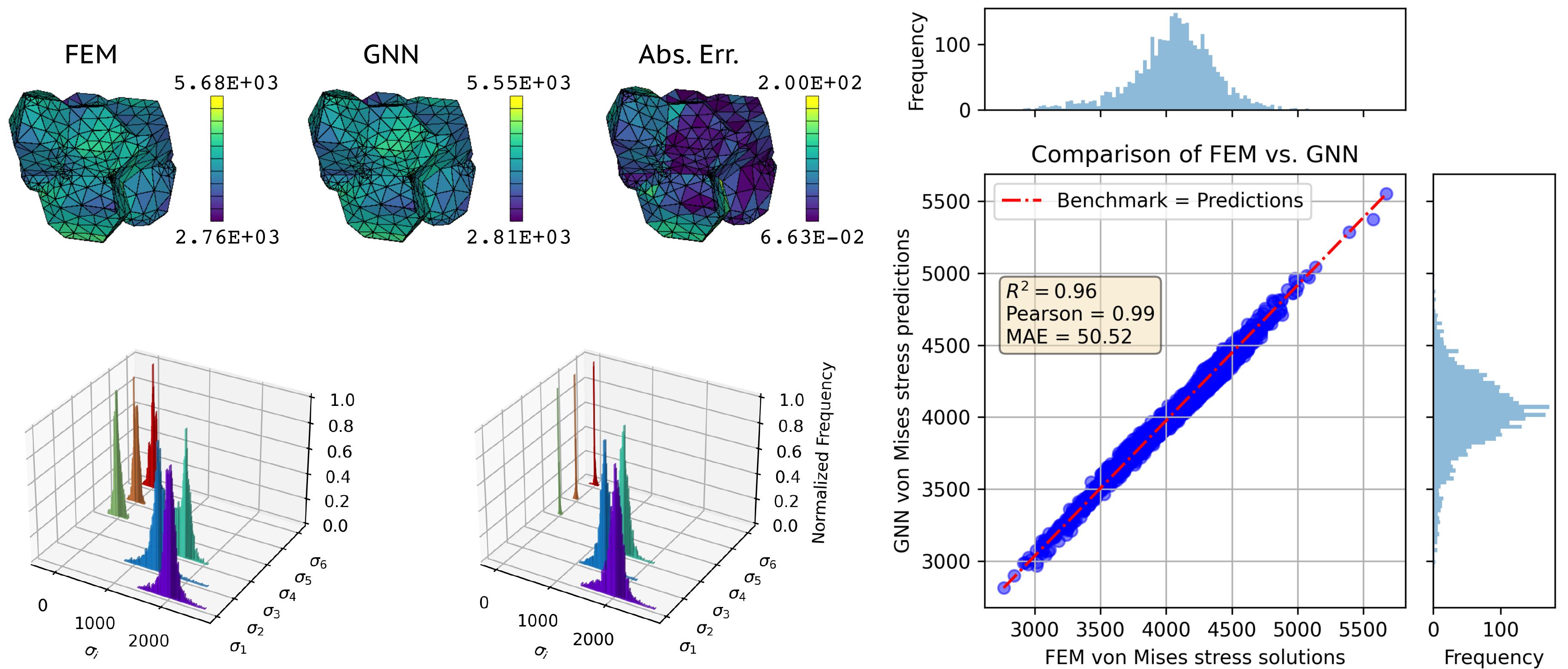}
    \caption{The evaluation of the prediction quality on von Mises stress for another example polycrystal. The upper left figures visualize the comparison between FEM and GNN predicted von Mises stress and absolute errors (visualized on elements with colorbars marked). The bottom left figures visualize the distributions of different stress components. The right figure shows the direct map between FEM and GNN predicted von Mises stresses.}
    \label{fig 10 test 3}
\end{figure}

Following the analysis procedure, we demonstrate the effective learning of the GNN by analyzing two other polycrystals with overall $R^2$ values of 0.992 and 0.991 from inferences, respectively (Figure \ref{fig 9 test 2} \& \ref{fig 10 test 3}). From the upper left figures, one observes very similar von Mises ranges are predicted by FEM and GNN, accompanied by low absolute errors. The quantitative comparison in the right figures confirms the qualitative observation for the von Mises stress inferences, with $R^2$ and Pearson coefficients of 0.96 and 0.99 for both the two polycrystals, respectively. Also, the two methods both predict similar stress component ranges demonstrated in the bottom left figures, illustrated by different color histograms. To summarize, these results demonstrate a few aspects of the robustness of the proposed GNN plasticity modeling: (1) overall stress components are predicted well by the high $R^2$ values, with no overfitting for testing sets; (2) general trends of stress components are captured; (3) von Mises stress are well learned verified both qualitatively and quantitatively, demonstrated via similar stress distribution and high $R^2$ and Pearson coefficients. Specifically, von Mises is not introduced (or constrained) in the training process. Additionally, the plasticity model is effectively learned from a limited amount of training data.

Figure \ref{fig 10 test 3} also reflects the potential limitation of the proposed method: the distribution of the stress components $\sigma_4$, $\sigma_5$, and $\sigma_6$ are not fully captured by the GNN. Qualitatively, one may argue that the variance of the data distribution around zero is not learned via GNN. This can be explained by the low-stress value range for the related stress components uncoupled with the loading direction (i.e., 1-direction). However, as can be visually observed and with Eqn.~\eqref{vonMises_eqn}, the stresses in the uncoupled directions do not significantly contribute to the von Mises stresses, considering the high-quality predictions on the von Mises stresses (Figs.~\ref{fig 9 test 2} \& \ref{fig 10 test 3}). The stress components correlated to the loading direction match well with the benchmark as illustrated in the left-bottom figure. 


\begin{figure}[htbp]
    \centering
    \includegraphics[width=.9\linewidth]{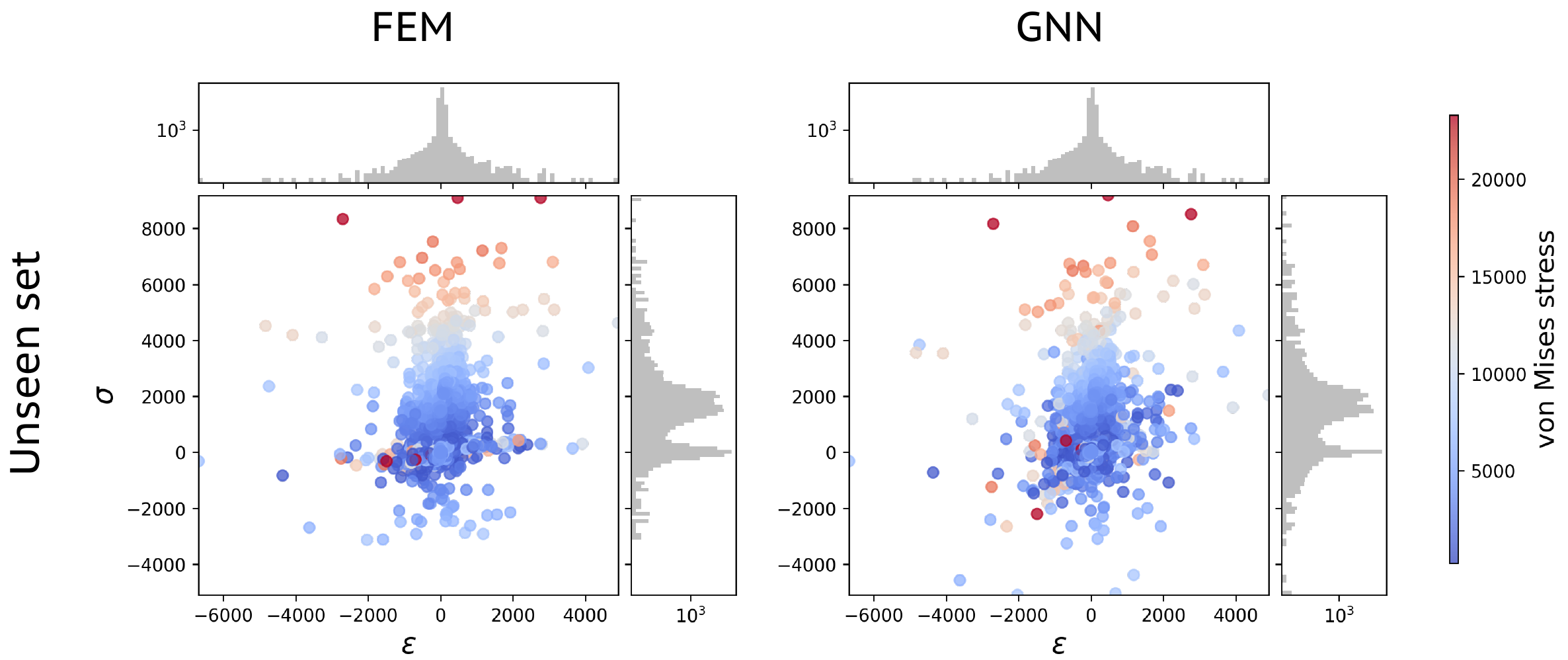}
    \caption{The comparison between the ground truth (by {\em FEPX}) and predictions on the strain-stress maps for all stress-strain components. The data points are visualized according to the calculated von Mises stresses.}
    \label{fig 12 stress ele compare}
\end{figure}

Figure~\ref{fig 12 stress ele compare} shows the comparison of the stress components predictions per cell for the overall $\epsilon \mapsto \sigma$ and $\epsilon_{11} \mapsto\sigma_{11}$ maps, respectively. The GNN inferences effectively preserve the stress data trends, including both the data distribution and the von Mises stress values. Interestingly, one may discern qualitatively higher discrepancies between the two methods in the ``low-stress regime.'' This observation aligns with the discussion on the limitations highlighted in Figure \ref{fig 10 test 3}: the values for stress components around zero are not accurately captured. Nonetheless, the model demonstrates high performance and provides good overall predictions on the strain-stress maps.

\begin{figure}[htbp]
    \centering
    \includegraphics[width=.5\linewidth]{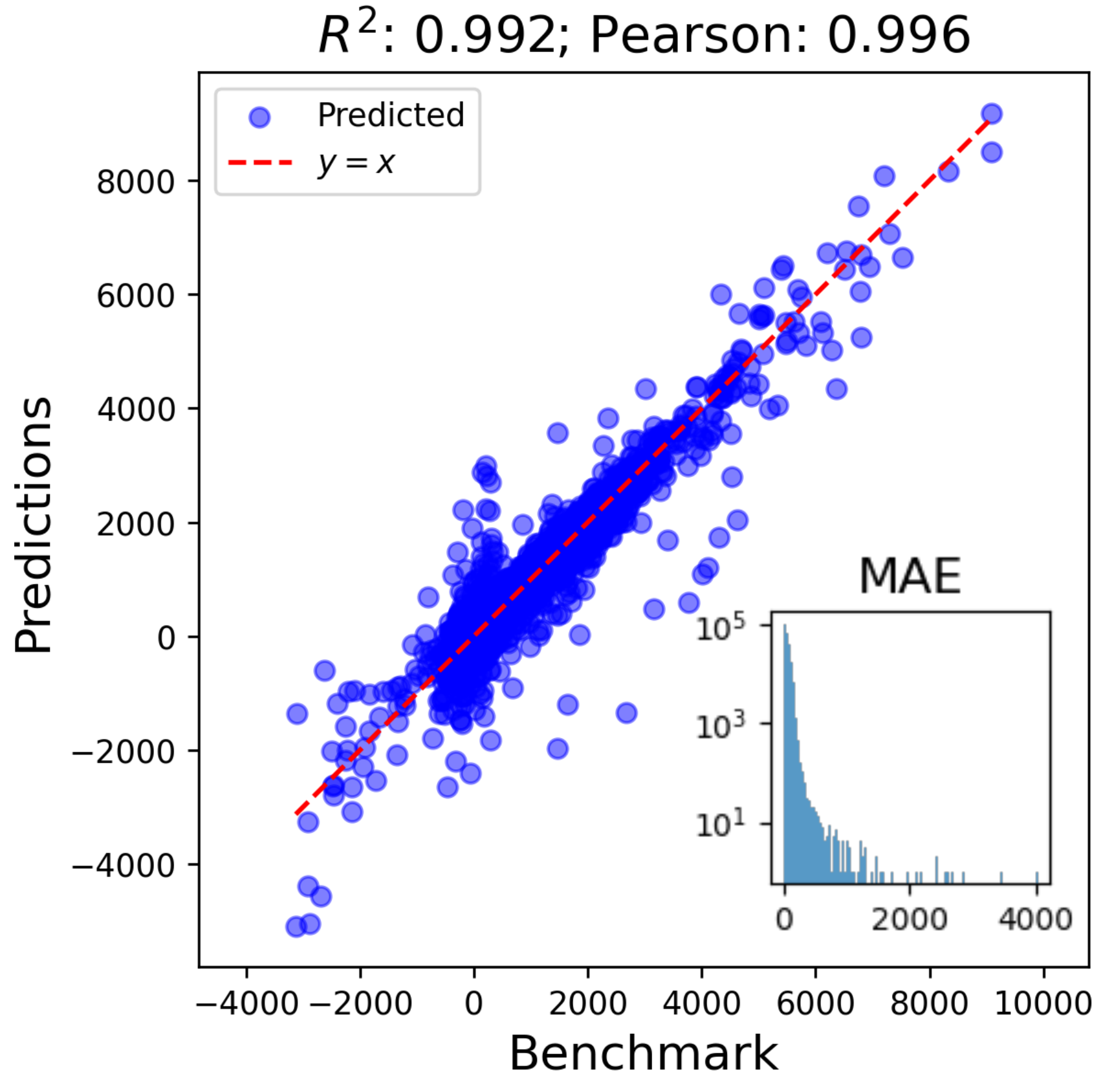}
    \caption{The overall prediction results on the validation dataset and the absolute error distribution.}
    \label{fig 11 unseen test}
\end{figure}

Figure \ref{fig 11 unseen test} presents the overall predictions of stress components for the 30 unseen simulations, demonstrating accurate predictions with an $R^2$ value of 0.992 and a Pearson coefficient of 0.996. These results indicate that the proposed GNN method not only generalizes well within the provided training and testing sets (i.e., interpolation) but also effectively extrapolates to data outside the given data regime.

\subsection{Discussions on loading-coupled directions\label{appendix stress loading}}

\begin{figure}[htbp]
    \centering
    \includegraphics[width=.9\linewidth]{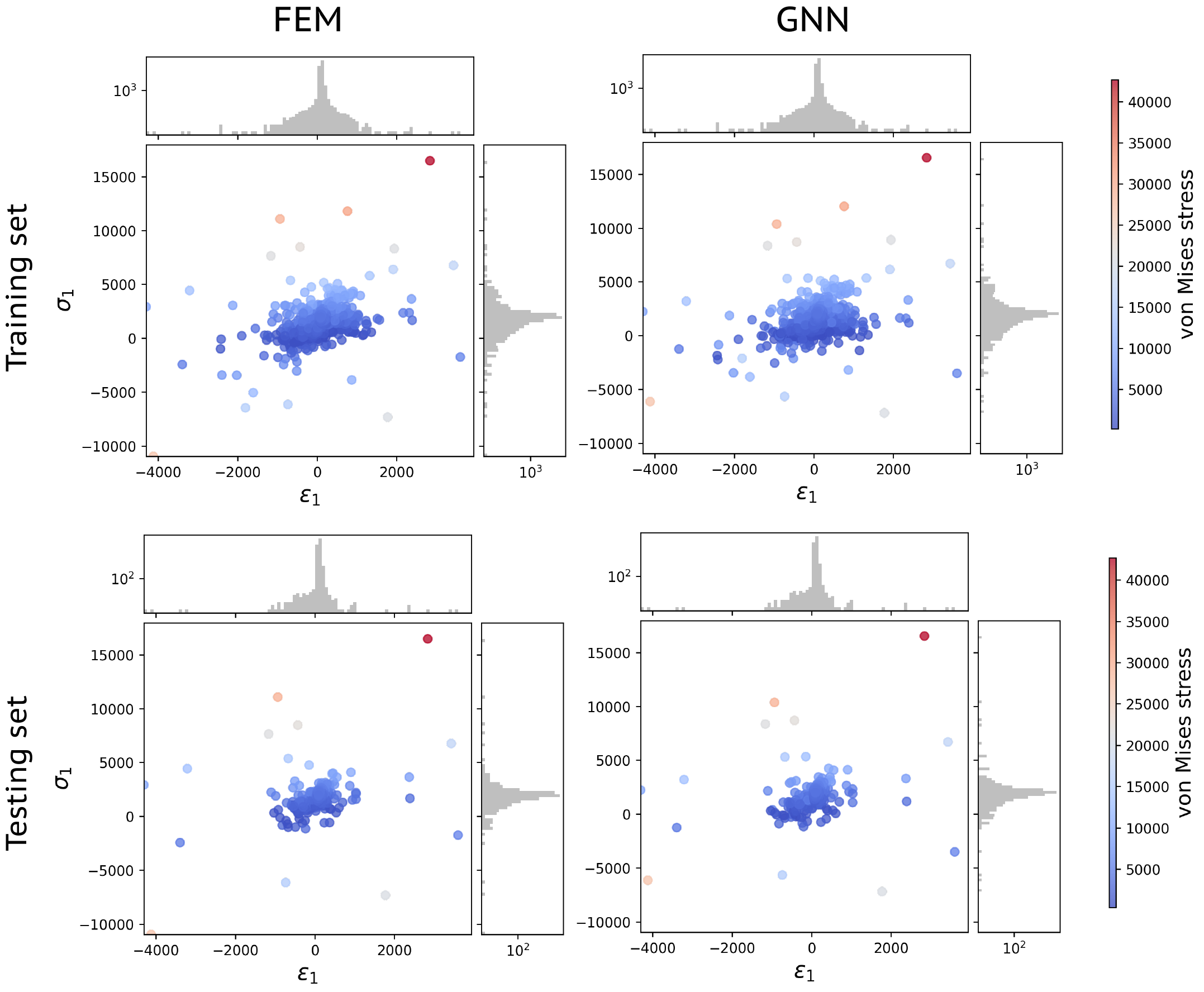}
    \caption{Comparison on FEM and GNN predictions on the $\epsilon_1$-$\sigma_1$ mapping between the training and testing sets corresponding to Figure \ref{fig 3 train & test}.}
    \label{fig dir 11 traintest}
\end{figure}
\begin{figure}[htbp]
    \centering
    \includegraphics[width=.9\linewidth]{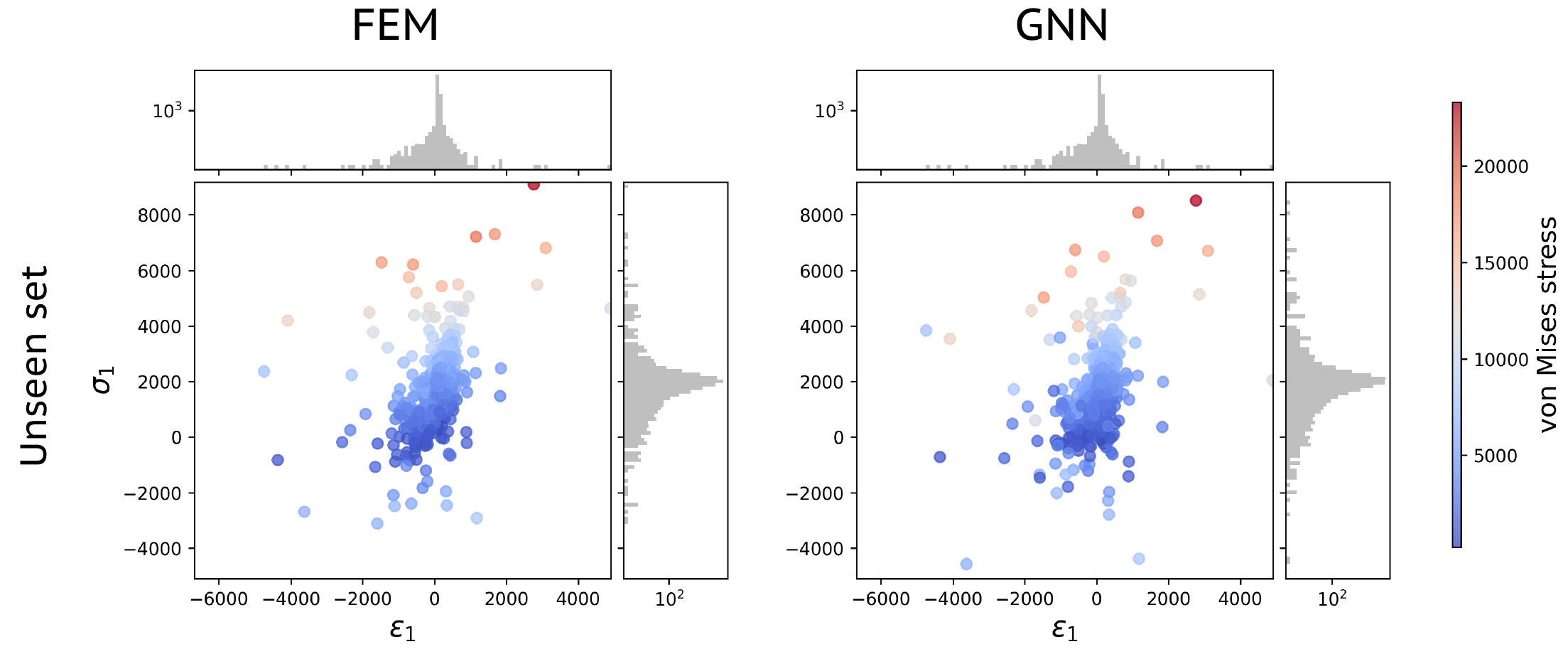}
    \caption{Comparison on FEM and GNN predictions on the $\epsilon_1$-$\sigma_1$ mapping for the validation sets corresponding to Figure \ref{fig 12 stress ele compare}.}
    \label{fig dir 11 unseen}
\end{figure}

Figures \ref{fig dir 11 traintest} and \ref{fig dir 11 unseen} illustrate the 1-directional strain-to-stress mapping, comparing FEM and GNN predictions alongside Figures \ref{fig 5 epsilon & sigma11} and \ref{fig 12 stress ele compare}. The results show that the GNN achieves significantly more accurate predictions for these maps. We hypothesize that this improved accuracy arises from the larger value range in the loading direction (and its coupled components), which may enhance the model's ability to predict the map more effectively.


\begin{figure}[htbp]
    \centering
    \includegraphics[width=.8\linewidth]{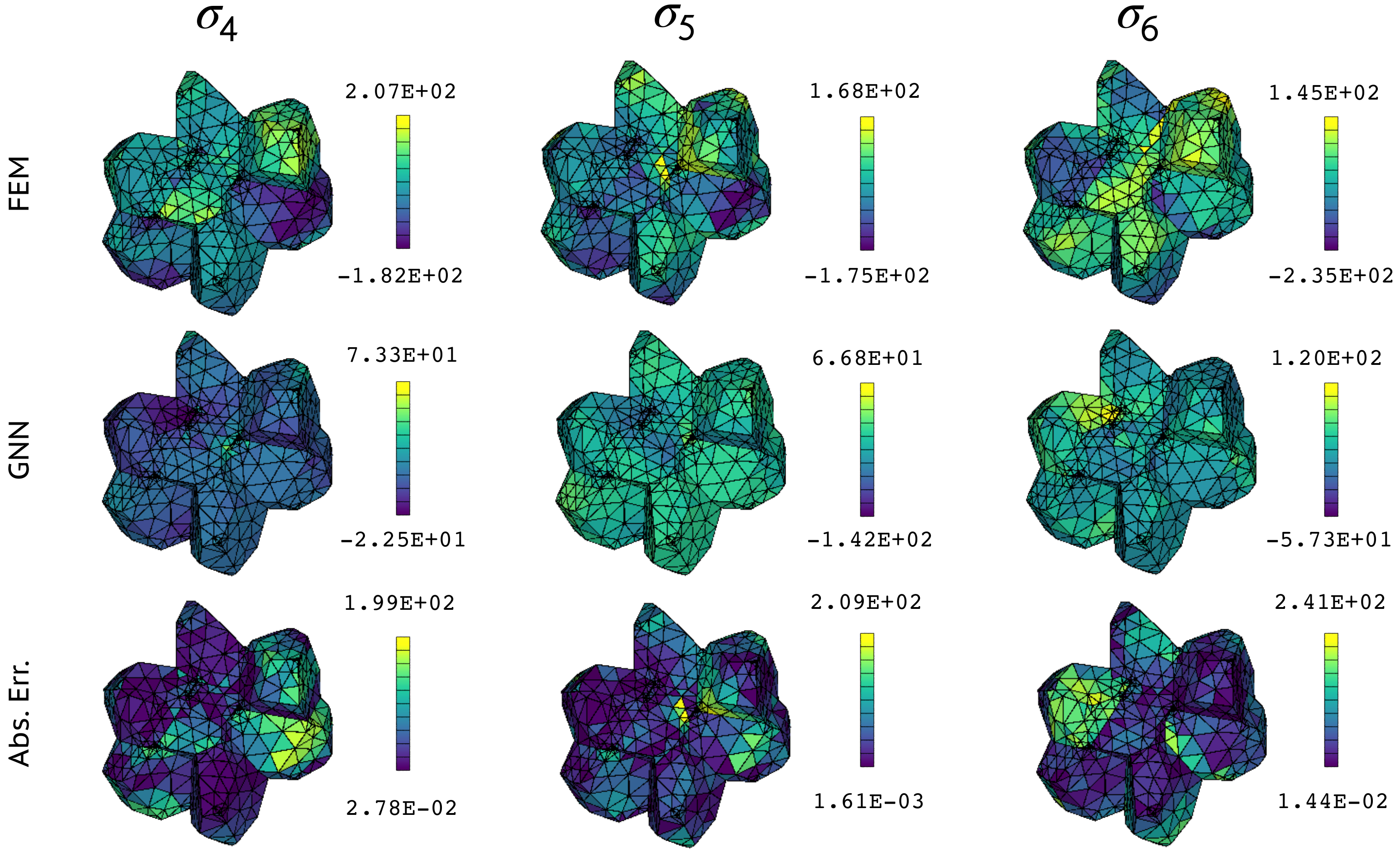}
    \caption{The comparison between FEM and GNN predictions for the loading uncoupled directions, with absolute errors of stress components on one example grain in the testing sets for stress components $\sigma_4$, $\sigma_5$ and $\sigma_6$ (visualized on elements with marked color bars). The unit for stress is [MPa].}
    \label{fig sig 345}
\end{figure}

To further elaborate, Figure \ref{fig sig 345} presents the predicted stress components $\sigma_3$, $\sigma_4$, and $\sigma_5$ (uncoupled from the loading direction), corresponding to Figure \ref{fig 6 stress components}. The predictions for these loading-direction-uncoupled components are observably less accurate. This visualization supports and reinforces our earlier discussions.

\subsection{Error distribution for stress components\label{appendix err stress}}

\begin{figure}[htbp]
    \centering
    \includegraphics[width=.9\linewidth]{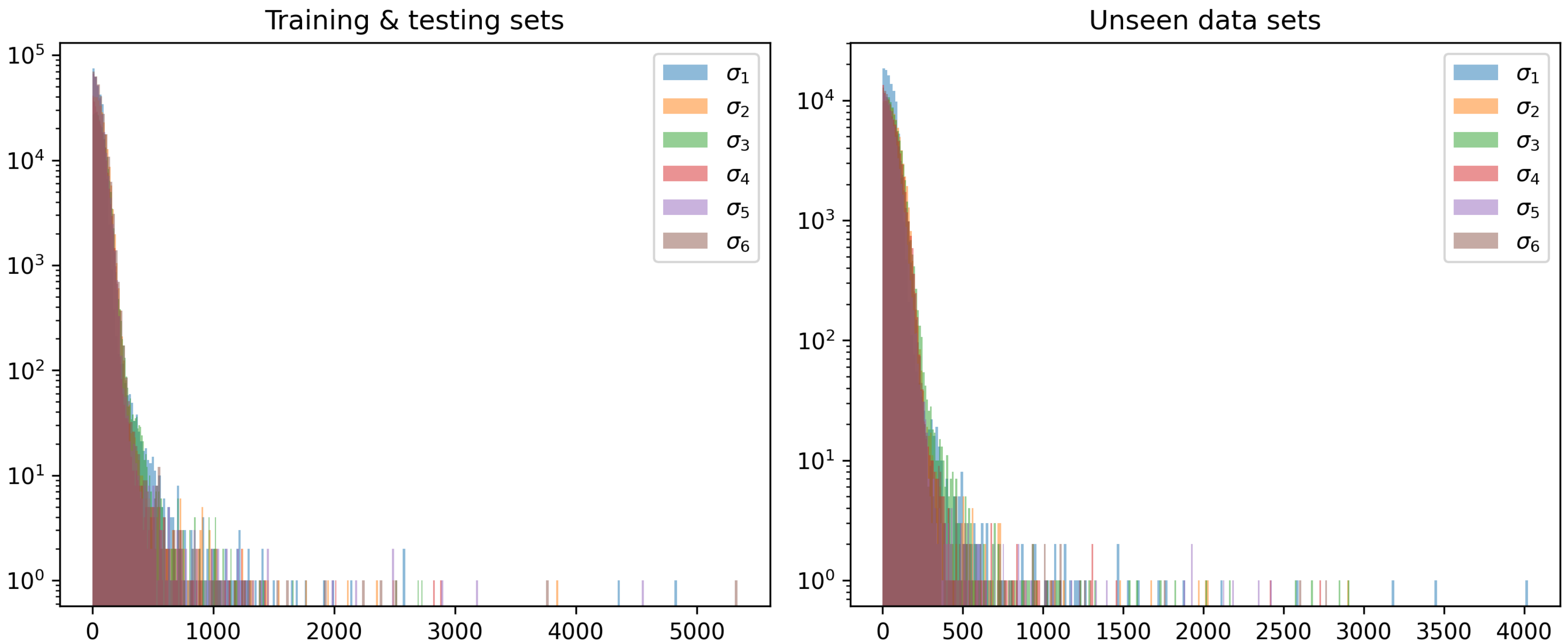}
    \caption{Error distribution between the training \& testing sets and the validation dataset (marked as unseen) for comparing the six stress components.}
    \label{fig err}
\end{figure}

Figure \ref{fig err} visualizes the prediction error distribution corresponding to Figure \ref{fig 15 error analysis}, comparing the error distributions across different stress components. The results confirm that the errors remain consistent across the training, testing, and validation datasets. Additionally, the data range shows minimal variation between the training and testing sets and the validation set. This serves as further evidence supporting Figure \ref{fig 15 error analysis}, demonstrating the strong generalizability of the GNN.

\clearpage
\bibliographystyle{elsarticle-num-names} 
\bibliography{cas-refs}

\end{document}